\documentclass[11pt]{article}

\usepackage{html,amsmath,amssymb,amsfonts,epsfig,cancel,cite,longtable,caption,siunitx,array,color,booktabs,setspace}

\usepackage{dsfont}
\usepackage{subfig}
\usepackage{braket}
\usepackage{appendix}
\usepackage{float}
\usepackage{tikz}
\usepackage{longtable}

\numberwithin{equation}{section}

\newcommand{\be}{\begin{equation}}
\newcommand{\ee}{\end{equation}}
\newcommand{\beq}{\begin{equation}}
\newcommand{\eeq}{\end{equation}}
\newcommand{\ba}{\begin{array}}
\newcommand{\ea}{\end{array}}
\newcommand{\bea}{\begin{eqnarray}}
\newcommand{\eea}{\end{eqnarray}}
\newcommand{\bean}{\begin{eqnarray*}}
\newcommand{\eean}{\end{eqnarray*}}

\def\p{\pi}

\def\gsim{ \lower .75ex \hbox{$\sim$} \llap{\raise .27ex \hbox{$>$}} }
\def\lsim{ \lower .75ex \hbox{$\sim$} \llap{\raise .27ex \hbox{$<$}} }
\def\be{\begin{equation}}
\def\ee{\end{equation}}
\def\bea{\begin{eqnarray}}
\def\eea{\end{eqnarray}}

\newcommand{\cN}{{\cal N}}

\newcommand{\cA}{{\cal A}}
\newcommand{\cB}{{\cal B}}
\newcommand{\cC}{{\cal C}}

\newcommand{\cV}{{\cal V}}

\newcommand{\fn}{\footnotesize}

\def\cjn1{{\cA, \cC^*\otimes \wedge^j \cN^*}}
\def\bjn1{{\cA, \cB^*\otimes \wedge^j \cN^*}}
\def\vjn1{{\cA, \cV^*\otimes \wedge^j \cN^*}}
\def\cjn2{{\cA, \cC\otimes \wedge^j \cN^*}}
\def\bjn2{{\cA, \cB\otimes \wedge^j \cN^*}}
\def\vjn2{{\cA, \cV\otimes \wedge^j \cN^*}}

\newcommand{\overbar}[1]{\mkern 3.0mu\overline{\mkern-4.0mu#1\mkern-0.5mu}\mkern 1.5mu}

\renewcommand{\footnoterule}{%
  \kern -3pt
  \hrule width 0.4\textwidth height 0.3pt
  \kern 3pt
}

\begin{document}

\title{{\LARGE \bf$~$\\[-7pt]
Particle Physics Model Building\\ with\\ Reinforcement Learning\\ [9pt]
}}

\vspace{3cm}

\author{
T.~R.~Harvey${}^{1}$ and
A.~Lukas${}^{1}$,
}
\date{}
\maketitle
\thispagestyle{empty}
\begin{center} { ${}^1${\it Rudolf Peierls Centre for Theoretical Physics, Oxford University,\\
       1 Keble Road, Oxford, OX1 3NP, U.K.}}
\end{center}

\vspace{11pt}
\abstract
\noindent
In this paper, we apply reinforcement learning to particle physics model building. As an example environment, we use the space of Froggatt-Nielsen type models for quark masses. Using a basic policy-based algorithm we show that neural networks can be successfully trained to construct Froggatt-Nielsen models which are consistent with the observed quark masses and mixing. The trained policy networks lead from random to phenomenologically acceptable models for over 90\% of episodes and after an average episode length of about 20 steps. We also show that the networks are capable of finding models proposed in the literature when starting at nearby configurations.
\vskip 5cm
{\hbox to 7cm{\hrulefill}}
\noindent{\fn thomas.harvey@physics.ox.ac.uk}\\
{\fn andre.lukas@physics.ox.ac.uk}
\newpage
\tableofcontents
\section{Introduction}
Machine learning in particle and string theory has developed into a fruitful and growing area of interdisciplinary research, triggered by the work in Refs.~\cite{He:2017Set,Ruehle:2017mzq}. (For a review and a comprehensive list of references see Ref.~\cite{RUEHLE20201}.)  Much of the activity to date has been in the context of supervised learning (see, for example, Refs.~\cite{Klaewer:2018sfl,Bull:2018uow,Wang:2018rkk,Brodie:2019dfx,Day:2019ucy,Krippendorf:2020gny,Gukov:2020qaj,Anderson:2020hux}), where data sets which arise in physics or related areas of mathematics have been used to train neural networks. However, there has also been some interesting work using reinforcement learning (RL), particular in relation to string model building~\cite{Halverson:2019tkf,Larfors:2020ugo}.\\[2mm]
In the present paper, we are interested in reinforcement learning in the context of particle physics model building. More precisely, we would like to address the following question. Can techniques of reinforcement learning be used to train a neural network to construct particle physics models with certain prescribed properties? At its most ambitious, such a network might be used to explore large classes of quantum field theories in view of their consistency with experimental data, thereby facilitating the search for physical theories beyond the standard model of particle physics. However, such a wide-ranging approach would require considerably conceptual work as well as computing resources and does not seem feasible for a first exploration. (For a different  approach to quantum field theory via methods of machine learning see Ref.~\cite{Halverson:2020trp}.)\\[2mm]
For this reason, we will focus on a much more limited arena of particle physics model building which can be relatively easily described and where extracting relevant physics properties is straightforward. Specifically, we will consider Froggatt-Nielsen (FN) models of fermion masses~\cite{Froggatt:1978nt,MassMatrix1,MassMatrixTwo,Dudas:1995yu,Dudas:1996fe,Ibanez:1994ig}, focusing on the quark sector. (For related early work on mass model building with horizontal $U(1)$ symmetries see also Refs.~\cite{Davidson:1983fy,Davidson:1987tr,Davidson:1989bx,Davidson:1979wr,Davidson:1981zd}.)\\[2mm] 
The standard model of particle physics contains the up and down quark Yukawa couplings  $Y^u_{ij}$ and $Y^d_{ij}$, where $i,j,\cdots=1,2,3$ label the three families. Within the standard model, these couplings are mere parameters inserted ``by hand". Upon diagonalisation, they determine the masses $(m_{u,i})=(m_u,m_c,m_t)$ and $(m_{d,i})=(m_d,m_s,m_b)$ of the up and down type quarks as well as the CKM mixing matrix $V_{\rm CKM}$.\\[2mm] 
FN models attempt to explain the values of $Y^u_{ij}$ and $Y^d_{ij}$ by introducing  $U_a(1)$ symmetries, where $a=1,\ldots ,r$, and singlet fields $\phi_\alpha$, where $\alpha=1,\ldots ,\nu$, in addition to the structure present in the standard model. The idea is that the Yukawa couplings are either zero, if forbidden by the $U_a(1)$ symmetries, or given in terms of the vacuum expectation values (VEVs) $\langle\phi\rangle$ of the scalar fields, such that $Y^u_{ij}\sim\langle\phi\rangle^{n_{ij}}$ and $Y^d_{ij}\sim\langle\phi\rangle^{m_{ij}}$. Here, $n_{ij}$ and $m_{ij}$ are (non-negative) integers whose values are determined by $U_a(1)$ invariance of the associated operator. A FN model is easily described by its charge matrix $({\mathcal{Q}_{a}}^I)=(q_a(Q^i),q_a(u^i),q_a(d^i),q_a(H),q_a(\phi))$, where $q_a$ denotes the charge with respect to $U_a(1)$, $Q_i$ are the left-handed quark-doublets, $u_i$ and $d_i$ are the right-handed up and down quarks and $H$ is the Higgs doublet. (As we will discuss, the VEVs $\langle\phi_\alpha\rangle$, which may also be considered as part of the definition of a FN models, will be fixed to certain optimal values for a given charge assignment.) We can, therefore, think of the space of FN models as the space of charge matrices $\mathcal{Q}$. For practical reasons, we will impose limits, $q_{\rm min}\leq \mathcal{Q}_{aI}\leq q_{\rm max}$, on the entries of this matrix, so that the space of models becomes finite. However, note that, even for one $U(1)$ symmetry ($r=1$), one singlet ($\nu=1$) and a modest charge range $-q_{\rm min}=q_{\rm max}=9$ we have of the order of $10^{13}$ models. This is quite sizeable, even though it is small compared to typical model numbers which arise in string theory.\\[2mm]
The idea of RL is to train a neural network with data obtained by exploring an environment, subject to a goal defined by a reward function. (See, for example, Ref.~\cite{sutton2018reinforcement} for an introduction.) It has been shown that RL can lead to impressive performance, even for very large environments, where systematic scanning is impossible~\cite{go}. It is, therefore, natural to ask whether RL can help explore the large model environments realised by quantum field theory and string theory. In the present paper, we will use RL to explore the space of FN models for the quark sector. More specifically, our environment consists of the set $\{\mathcal Q\}$ of all FN charge matrices for a given number, $r$, of $U(1)$ symmetries, a given number, $\nu$, of singlets $\phi_\alpha$ and charges constrained by  $q_{\rm min}\leq \mathcal{Q}_{aI}\leq q_{\rm max}$. An action within this environment simply amounts to adding or decreasing one of the charges ${\mathcal{Q}_{a}}^I$ by one and a reward is computed based on how well the models reproduce the experimental quark masses and mixings. A terminal state is one that reproduces the experimental masses and mixing to a given degree of accuracy.

We use a simple policy-based RL algorithm, with a single policy network whose input is, essentially, the charge matrix $\mathcal{Q}$ and whose output is an action. The hope is that a successfully trained policy network of this kind will produce episodes starting from arbitrary (and typically physically unacceptable) FN models and efficiently lead to phenomenologically viable FN models.\\[2mm]
The plan of the paper is as follows. In the next section, we briefly review the theoretical background of this work, namely RL and FN model building, mainly to set the scene and fix notation. In Section~\ref{sec:setup} we describe our RL set-up and Section~\ref{sec:Training} presents the results we obtained for the cases of one singlet and one $U(1)$ symmetry and two singlets and two $U(1)$ symmetries. The appendices contain  a number of interesting FN models found by the neural network. 

\section{Background}\label{sec:Background}
\subsection{Reinforcement Learning}\label{sec:Reinforce}
We start with a quick overview of RL,  focusing on the aspects needed for this paper. For a comprehensive review see, for example, Refs.~\cite{sutton2018reinforcement} and \cite{RUEHLE20201}.\\[2mm]
The main components of an RL system are the {\it environment}, the {\it agents} and the {\it neural network(s)}. The latter are set up to learn certain properties of the environment, based on data delivered as the agent explores the environment. The mathematical underpinning of RL is provided by a {\it Markov decision process} (MDP), defined as a tuple $(\mathcal{S},\mathcal{A},\mathcal{P},\gamma,\mathcal{R})$. Here $\mathcal{S}$ is a set which contains the environment's states, $\mathcal{A}$ is a set of maps $\alpha:\mathcal{S}\rightarrow\mathcal{S}$ which represent the actions, $\mathcal{P}$ provides a probability $\mathbb{P}(S=s'|S=s,A=\alpha)$ for a transition from state $s$ to state $s'$ via the action $\alpha$, $\gamma\in [0,1]$ is called the {\it discount factor} and $\mathcal{R}:\mathcal{S}\times\mathcal{A}\rightarrow\mathbb{R}$ is the {\it reward function}. Among the states in $\mathcal{S}$  a subset of so-called {\it terminal states} is singled out which may, for example, consist of states with certain desirable properties. Within this set-up we can consider a sequence 
\[
 s_0\stackrel{\alpha_0,r_0}{\xrightarrow{\hspace*{8mm}}} s_1\stackrel{\alpha_1,r_1}{\xrightarrow{\hspace*{8mm}}}s_2\stackrel{\alpha_2,r_2}{\xrightarrow{\hspace*{8mm}}}s_3\cdots
\]
of states $s_t$ and actions $\alpha_t$, producing rewards $r_t$, where $t=0,1,2,\cdots$, which is referred to as an {\it episode}. In principle, an episode can have infinite length, although in practice a finite maximal episode length, $N_{\rm ep}$, is imposed. If an episode arrives at a terminal state before it reaches its maximal number of steps it is stopped. The return, $G_t$, of a state $s_t$ in such an episode is defined as
\begin{equation}
 G_t=\sum_{k\geq 0} \gamma^k r_{t+k}\; .
\end{equation} 
The discount factor $\gamma$ can be dialled to small values in order to favour short-term rewards dominating the return, or to values close to one so that longer-term rewards affect the return as well. The choice of action in a MDP is guided by a {\it policy} $\pi$, which provides probabilities $\pi(\alpha|s)=\mathbb{P}(A_t=\alpha|S_t=s)$ for applying a certain action $\alpha$ to a state $s$. Relative to such a policy, two important value functions, namely the {\it state value function} $V_\pi$ and the {\it state-action value function} $Q_\pi$, can be defined as expectation values of the return.
\begin{equation}
 V_\pi (s)=\mathbb{E}(G_t|S_t=s)\;,\qquad Q_\pi(s,\alpha)=\mathbb{E}(G_t|S_t=s,A_t=\alpha)\; .
\end{equation} 
The purpose of an RL system is to maximise a value function (state or state-action) over the set of possible policies. In practice, this can be realised in a number of ways which differ by which of the functions $\pi$, $V_\p$ and $Q_\pi$ are represented by neural networks and how precisely these neural networks are trained via exploration of the environment. Common to all algorithms is an iterative approach, where a batch of data, in the form of triplets $(s_t,a_t,G_t)$, is collected from episodes which are guided by the neural network(s) in their present state. This data is then used to update the neural network(s), followed by a further round of exploration and so on.\\[2mm]
For our purposes, we will consider what is probably the simplest approach, a basic policy-based algorithm referred to as REINFORCE. This set-up contains a single neural network $\pi_\theta$ with weights $\theta$ which represents the policy $\pi$. Its inputs are states and the outputs are probabilities for actions. Exploration of the environment is guided by the policy, meaning the steps in an episode are selected based on $\pi_\theta$, so
\begin{equation}\label{eppi}
 s_0\stackrel{\pi_\theta}{\xrightarrow{\hspace*{8mm}}} s_1\stackrel{\pi_\theta}{\xrightarrow{\hspace*{8mm}}}s_2\stackrel{\pi_\theta}{\xrightarrow{\hspace*{8mm}}}s_3\cdots \; .
\end{equation}
Data is collected by performing such episodes successively, so we can say that the system contains only one agent. According to the {\it policy-gradient theorem}, the neural network $\pi_\theta$ should be trained on the loss function
\begin{equation}\label{loss}
 L(\theta)=Q_\pi(s,a)\ln (\pi_\theta(s,a))\; ,
\end{equation} 
where $Q_\pi(s,a)$ can, in practice, be replaced by the return $G$ of the state $s$. Schematically, the algorithm then proceeds as follows.
\begin{enumerate}
\item[(1)] Initialise the policy network $\pi_\theta$.
\item[(2)] Collect a batch of data triplets $(s_t,a_t,G_t)$  from as many episodes~\eqref{eppi} as required.
New episodes start at random states $s_0$.
\item[(3)] Use this batch to update the weights $\theta$ of the policy network $\pi_\theta$, based on the loss~\eqref{loss}.
\item[(4)] Repeat from (2) until the loss is sufficiently small so that the policy has converged.
\end{enumerate}
 
\subsection{Froggatt-Nielsen models}\label{sec:FNMech}
Before we discuss Froggatt-Nielsen models, we quickly review fermion masses in the standard model of particle physics, in order to set up notation and present the experimental data.\\[2mm]
The standard model contains Yukawa interactions, which are responsible for generating the masses and mixing of quarks and leptons. In this paper, we focus on the quark sector for simplicity, although we expect that our work can be generalised to include the lepton sector. The quark Yukawa couplings in the standard model take the form
 \begin{equation}\label{smyuk}
     \mathcal{L}_{\rm Yuk}=Y^u_{ij} \overbar{Q}^i H^c u^j + Y^d_{ij} \overbar{Q}^i H d^j + \mbox{h.c.}\;,
\end{equation}
where $Q_i$ are the left-handed quarks, $u_i$, $d_i$ are the right-handed up and down type quarks and $H$ is the Higgs doublet. We use indices $i,j,\ldots=1,2,3$ to label the three families. Within the standard model, the Yukawa matrices $Y^u$ and $Y^d$ are not subject to any theoretical constraints - their (generally complex) values are inserted ``by hand" in order to fit the experimental results for masses and mixing. 

Once the charge-neutral component $H^0$ in the Higgs doublet develops a VEV, $v=\langle H^0\rangle$, the above Yukawa terms lead to Dirac mass terms with associated mass matrices
 \begin{equation}\label{MY}
     M_u = v\, Y^u\;,\qquad M_d = v\, Y^d\; .
\end{equation}
These matrices need to be diagonalised,
\begin{equation}\label{diag}
 M_u = U_u \hat{M}_u V_u^\dagger\;,\quad M_d = U_d \hat{M}_d V_d^\dagger\;,\quad\mbox{where}\quad  \hat{M}_u = \text{diag}(m_u,m_c,m_t) \;,\quad  \hat{M}_d = \text{diag}(m_d,m_s,m_b)\; ,
\end{equation} 
by unitary matrices $U_u$, $V_u$, $U_d$, $V_d$ in order to obtain the masses $(m_u,m_c,m_t)$ for the up-type quarks and the masses $(m_d,m_s,m_b)$ for the down-type quarks. The other observable quantity obtained from the quark Yukawa couplings is the Cabibbo-Kobayashi-Maskawa (CKM) matrix $V_{CKM}$, defined by
\begin{equation}\label{VCKM}
        V_{CKM} = U_u^\dagger U_d = 
        \left(
            \begin{array}{ccc}
             c_{12}c_{13} & s_{12}c_{13} & s_{13}e^{-i\delta} \\
             -s_{12}c_{23}-c_{12}s_{23}s_{13}e^{i\delta} & c_{12}c_{23}-s_{12}s_{23}s_{13}e^{i\delta} & s_{23}c_{13} \\
             s_{12}s_{23} -c_{12}c_{23}s_{13}e^{i\delta} & -c_{12}s_{23}-s_{12}c_{23}s_{13}e^{i\delta} & c_{23}c_{13} \\
            \end{array}
        \right).
\end{equation}
The CKM matrix is unitary and can, hence, be written in terms of three angles $\theta_{12},\theta_{13},\theta_{23}$ and a phase $\delta$, as in the above equation, where the abbreviations $s_{ij}=\sin(\theta_{ij})$ and $c_{ij}=\cos(\theta_{ij})$ have been used. The experimentally measured values for these quantities are given in Table~\ref{tab:exp}.
\begin{table}[!h]
\renewcommand{\arraystretch}{1.25}
\centering
   \begin{tabular}{|c|c|c|c|c|c|}\hline
         $m_u$ & $m_d$ & $m_c$ & $m_s$ & $m_t$ & $m_b$\\\hline
         ${0.00216}^{+0.00049}_{-0.00026}$ & ${0.00467}^{+0.00048}_{-0.00017}$ & ${1.27}\pm{0.02}$ & ${0.093}^{+0.011}_{-0.005}$ & $172.4 \pm 0.07$ & ${4.18}^{+0.03}_{-0.02}$\\\hline\hline
          $v$ & $s_{12}$ & $s_{13}$ &  $s_{23}$ & $\delta$ & \\\hline
         $\sim 174$ & $0.22650 \pm 0.00048$ & $0.00361^{+0.00011}_{-0.00009}$ & $0.04053^{+0.00083}_{-0.00061}$ & $1.196^{+0.045}_{-0.043}$ & \\
         \hline
      \end{tabular}
  \caption{\sf Experimentally measured masses in GeV and mixing angles of quarks from Ref.~\cite{PDG}.} \label{tab:exp}
\end{table}
Inserting the mixing angles and the phase from Table~\ref{tab:exp} into the parametrisation~\eqref{VCKM} gives the numerical CKM matrix
\begin{equation}
        \label{eqn:CKMVal}
        \left| V_{CKM} \right| \approx \left(
            \begin{array}{rrr}
             0.9740 & 0.2265 & 0.0036 \\
             0.2264 & 0.9732 & 0.0405 \\
             0.0085 & 0.0398 & 0.9992 \\
            \end{array}
            \right).
\end{equation}
In the context of the standard model, the Yukawa matrices $Y^u$ and $Y^d$ in Eq.~\eqref{smyuk} have to be chosen to fit these experimental values for masses and mixing but this still leaves considerable freedom. Only $10$ real constrains are imposed on the $36$ real parameters which determine $Y^u$ and $Y^d$.\\[2mm]
Froggatt-Nielsen (FN) models provide a framework for adding more structure to the Yukawa sector of the standard model, in an attempt to remove some of this ambiguity and provide a theoretical explanation for the observed masses and mixing. Two main ingredients are added to the picture: a number of global $U(1$) symmetries $U_a(1)$, where $a=1,\ldots ,r$, and a number of complex scalar fields $\phi_\alpha$, where $\alpha=1,\ldots ,\nu$, which are singlets under the standard model gauge group. The standard model fields as well as the scalar singlets are assigned $U_a(1)$ charges which we denote by $q_a(Q^i)$, $q_a(u^i)$, $q_a(d^i)$, $q_a(H)$ and $q_a(\phi_\alpha)$. In fact, to simplify matters, we assume that we have the same number of $U(1)$ symmetries and singlet fields, $\nu=r$, and that the $a^{\rm th}$ singlet $\phi^a$ is only charged under $U_a(1)$. The resulting singlet charges are then denoted by $q_a(\phi)$.\\[2mm]
Given this set-up, the standard model Yukawa couplings~\eqref{smyuk} are no longer in general consistent with the $U_a(1)$ symmetries and should be replaced by
\begin{equation}
 {\cal L}_{\rm Yuk} =\sum_{i,j}\left(a_{ij} \phi_1^{n_{1,ij}}\cdots \phi_r^{n_{r,ij}} \overbar{Q}^i H^c u^j\,+\,b_{ij} \phi_1^{m_{1,ij}}\cdots \phi_r^{m_{r,ij}} \overbar{Q}^i H d^j\right) + \mbox{h.c.}
\end{equation}
where $n_{a,ij}$ and $m_{a,ij}$ are non-negative integers. For a term $(ij)$ in the up-quark sector to be invariant under $U_a(1)$ we require the conditions
\begin{equation} \label{nij}
 n_{a,ij}=-\frac{q_a(\overbar{Q}^i H^c u^j)}{q_a(\phi)}\; .
\end{equation}
Hence, the term $(ij)$ in the u-quark sector is allowed if the $n_{a,ij}$ given by Eq.~\eqref{nij} are non-negative integers for all $a=1,\ldots ,r$. In this case, the coefficient $a_{ij}$ is of order one, otherwise it is set to zero. An analogous rule applies to the terms for the down-type quarks. Once the scalars $\phi_a$ develop VEVs, $v_a=\langle\phi_a\rangle$, Yukawa couplings
\begin{equation}\label{yukfn}
 Y^u_{ij}=a_{ij}v_1^{n_{1,ij}}\cdots v_r^{n_{r,ij}}\;,\qquad Y^d_{ij}=b_{ij}v_1^{m_{1,ij}}\cdots v_r^{m_{r,ij}}\; .
\end{equation}
are generated~\footnote{If these Yukawa couplings are generated at a high energy scale they have to be renormalised down to the electro-weak scale, in order to facilitate comparison with the experimental values. Since this typically leads to order one coefficients which have already been included via $a_{ij}$, $b_{ij}$ we will not consider this explicitly.}. The main model building idea in this setting is that moderately small singlet VEVs $v_a$ can generate the required large hierarchies in masses, in a way that is controlled by the integers $n_{a,ij}$ and $m_{a,ij}$ and, hence, ultimately, by the choices of $U_a(1)$ charges.\\[2mm]
At this stage the environment of FN models consists of $U_a(1)$ charges for all fields, the singlet VEVs $v_a$ and the coefficients $a_{ij}$, $b_{ij}$. In principle, the singlet VEVs are meant to be fixed by a scalar potential but implementing this in detail adds another layer of model building. Instead, for a given choice of charges and coefficients $a_{ij}$, $b_{ij}$, we will fix the VEVs $v_a$ such that the model provides an optimal fit to the experimental masses and mixing. The non-zero coefficients  $a_{ij}$, $b_{ij}$ might be considered as part of the environment definition but, to keep things simply, we will fix those to specific numerical values of order one. While, in general, $a_{ij}$ and $b_{ij}$ can be complex, we simplify this scenario by only allowing them to take real values. Consequently, we will not attempt to fit the CP violating phase $\delta$ in the CKM matrix. As a further simplification, we require that the top Yukawa term $\bar{Q}^3H^cu^3$ is present without any singlet insertions, a condition which seems reasonable given the size of the top Yukawa coupling. This requirement can be used to fixed the $U_a(1)$ charges of the Higgs multiplet as
\begin{equation}
 q_a(H)=q_a(u^3)-q_a(Q^3)\; .
\end{equation}
Altogether, this means a FN model within our set-up is specified by the charges choices
\begin{equation}\label{env}
\left( {\mathcal{Q}_a}^I\right)=\left(q_a(Q^i),q_a(u^i),q_a(d^i),q_a(\phi)\right)\; ,
\end{equation}
which we have assembled into the $r\times 10$ integer charge matrix $\mathcal{Q}$. In practice, the charges in $\mathcal{Q}$ will be restricted to a certain range
\begin{equation}\label{envrange}
 q_{\rm min}\leq {\mathcal{Q}_a}^i\leq q_{\rm max}\; ,
\end{equation}
with $q_{\rm min}$ and $q_{\rm max}$  to be specified later. While this leads to a finite space of charge matrices and associated FN models, numbers can be considerable. For example, for $-q_{\rm min}=q_{\rm max}=9$ we have $\sim 10^{13}$ models in the case of a single $U(1)$ symmetry and $\sim 10^{26}$ models for the case of two $U(1)$ symmetries.\\[2mm]
The environment~\eqref{env} of FN models has a number of permutation degeneracies, since the assignment of charges to families and the order of $U_a(1)$ symmetries does not carry physical meaning, although part of this symmetry is broken by designating $Y_{33}^u$ the top Yukawa coupling. This means there is a permutation degeneracy isomorphic to
 \begin{equation}\label{psymm}
  S_2\times S_2\times S_3\times S_r
 \end{equation} 
in the environment~\eqref{env}. For the purpose of RL we will not attempt to remove this redundancy, as this would complicate the constraints on the charges in $\mathcal{Q}$.\\[2mm]
From the viewpoint of particle physics model building the task is now to investigate the model landscape defined by Eq.~\eqref{env} and extract the phenomenologically promising cases. Considerable effort has been invested into this, since the original proposal of Froggatt and Nielsen~\cite{Froggatt:1978nt}. It is precisely this task we wish to carry out using reinforcement learning.
    
\section{Model building with reinforcement learning}\label{sec:setup}
We now explain how we propose to map the problem of FN model building onto the structure of reinforcement learning. We begin by describing the set-up of the RL environment.

\subsection{The environment}
We need to identify how the various ingredients of a MDP are realised in our context. We take the set $\mathcal{S}$ of states to consists of all FN models for a fixed number, $r$, of $U(1)$ symmetries and the same number of singlet fields. These models are represented by the $r\times 10$ integer charge matrices $\mathcal{Q}$ in Eq.~\eqref{env}, with entries restricted as in Eq.~\eqref{envrange}. The set $\mathcal{A}$ of actions $\alpha$ consists of the basic operations
\begin{equation}\label{action}
 {\mathcal{Q}_a}^I\stackrel{\alpha}{\longrightarrow} {\mathcal{Q}_a}^I\pm 1\; ,
\end{equation}
that is, increasing or decreasing a single charge ${\mathcal{Q}_a}^I$ by one while keeping all other charges unchanged. These are deterministic actions so we do not need to introduce transition probabilities $\mathcal{P}$. The number of different actions is $2\times r\times 10=20r$. For the discount factor $\gamma$ we choose the value $\gamma=0.98$.\\[2mm]
Defining the reward function $\mathcal{R}$ requires a bit more effort. We start by defining the intrinsic value for a state $\mathcal{Q}$ as
\begin{equation}\label{VQ}
 \mathcal{V}(\mathcal{Q})=-\min_{|v_a|\in I}\sum_\mu \left|{\rm log}_{10}\left(\frac{|\mu_{\mathcal{Q},v_a}|}{|\mu_{\rm exp}|}\right)\right|\; .
\end{equation} 
Here, $\mu$ runs over the six quark masses as well as the entries of the CKM matrix, $\mu_{\mathcal{Q},v_a}$ is the value for one of these quantities predicted by the model with charge matrix $\mathcal{Q}$ and scalar fields VEVs $v_a$, computed from Eqs.~\eqref{yukfn}, \eqref{MY}, \eqref{diag}, \eqref{VCKM} (using fixed random values of the order-one coefficients $a_{ij}$, $b_{ij}$), and $\mu_{\rm exp}$ is its experimental value as given in Table~\eqref{tab:exp} and Eq.~\eqref{eqn:CKMVal}. The minimisation is carried out over the scalar field VEVs $v_a$, in a certain range $I=[v_{\rm min},v_{\rm max}]$, with typical values $v_{\rm min}=0.01$ and $v_{\rm max}=0.3$. From this definition, the intrinsic value of a state $\mathcal{Q}$ is simply the (negative) total order of magnitude by which predicted masses and mixings deviate from the experimental ones, for optimal choices of the scalar field VEVs.

A terminal state $\mathcal{Q}$ in our environment is one which is phenomenologically promising, that is, a state which gives rise to (roughly) the correct masses and mixings. More specifically, we call a state terminal if its intrinsic value $\mathcal{V}(\mathcal{Q})$ is larger than a certain threshold value $\mathcal{V}_0$ and if each individual deviation $-|{\rm log}_{10}(|\mu_\mathcal{Q}|/|\mu_{\rm exp}|)|$ (computed for the scalar field VEVs which minimise Eq.~\eqref{VQ}) is larger than a threshold value $\mathcal{V}_1$. Since we have fixed our order-one parameters $a_{ij}$, $b_{ij}$ these threshold values are chosen relatively generously, so as to not miss any promising models. For our computations, we have used $\mathcal{V}_0=-10$ and $\mathcal{V}_1=-1$.\\[2mm]
Based on this intrinsic value, the reward $ \mathcal{R}(\mathcal{Q},\alpha)$ for an action $\mathcal{Q}\stackrel{\alpha}{\rightarrow}\mathcal{Q}'$ of the form~\eqref{action}, connecting two states $\mathcal{Q}$ and $\mathcal{Q}'$, is defined by
\begin{equation}\label{reward}
 \mathcal{R}(\mathcal{Q},\alpha)=\left\{\begin{array}{cll} \mathcal{V}(\mathcal{Q}')-\mathcal{V}(\mathcal{Q})&\mbox{if}& \mathcal{V}(\mathcal{Q}')-\mathcal{V}(\mathcal{Q})>0\\\mathcal{R}_{\rm offset}&\mbox{if}& \mathcal{V}(\mathcal{Q}')-\mathcal{V}(\mathcal{Q})\leq 0\end{array}\right.\; .
\end{equation}
Here, $\mathcal{R}_{\rm offset}$ is a fixed (negative) value which penalises a decrease of the intrinsic value, typically chosen as $\mathcal{R}_{\rm offset}=-10$. In addition, if the new state $\mathcal{Q}'$ is terminal a terminal bonus $\mathcal{R}_{\rm term}$, typically chosen as  $\mathcal{R}_{\rm term}=100$, is added to the reward~\eqref{reward}.

\subsection{Neural network}\label{sec:nn}
To represent the policy $\pi$, we use a fully connected network $f_\theta$ with the following structure.
\begin{center}
\begin{tikzpicture}
\draw[->,thick] (0.35,0) -- (0.7,0);
\draw[thick] (0.7,0.25) -- (0.7,-0.25) -- (1.9,-0.25) -- (1.9,0.25) -- (0.7,0.25);
\draw[->,thick] (1.9,0) -- (2.6,0);
\draw[thick] (2.6,0.25) -- (2.6,-0.25) -- (3.8,-0.25) -- (3.8,0.25) -- (2.6,0.25);
\draw[->,thick] (3.8,0) -- (4.5,0);
\draw[thick] (4.5,0.25) -- (4.5,-0.25) -- (5.7,-0.25) -- (5.7,0.25) -- (4.5,0.25);
\draw[->,thick] (5.7,0) -- (6.4,0);
\draw[thick] (6.4,0.25) -- (6.4,-0.25) -- (7.6,-0.25) -- (7.6,0.25) -- (6.4,0.25);
\draw[->,thick] (7.6,0) -- (8.3,0);
\draw[thick] (8.3,0.25) -- (8.3,-0.25) -- (9.5,-0.25) -- (9.5,0.25) -- (8.3,0.25);
\draw[->,thick] (9.5,0) -- (10.2,0);
\draw[thick] (10.2,0.25) -- (10.2,-0.25) -- (11.4,-0.25) -- (11.4,0.25) -- (10.2,0.25);
\draw[->,thick] (11.4,0) -- (12.1,0);
\draw[thick] (12.1,0.25) -- (12.1,-0.25) -- (13.2,-0.25) -- (13.2,0.25) -- (12.1,0.25);
\draw[->,thick] (13.2,0) -- (13.9,0);
\draw[thick] (13.9,0.25) -- (13.9,-0.25) -- (15.1,-0.25) -- (15.1,0.25) -- (13.9,0.25);
\draw[->,thick] (15.1,0) -- (15.45,0);
\node at (1.3,0.02) {\scriptsize affine};
\node at (3.2,0.02) {\scriptsize SELU};
\node at (5.1,0.02) {\scriptsize affine};
\node at (7,0.02) {\scriptsize SELU};
\node at (8.9,0.02) {\scriptsize affine};
\node at (10.8,0.02) {\scriptsize SELU};
\node at (12.7,0.02) {\scriptsize affine};
\node at (14.55,0.02) {\scriptsize softmax};
\node at (0,0.05) {\scriptsize $\mathbb{R}^{10r}$};
\node at (15.85,0.05) {\scriptsize $\mathbb{R}^{20r}$};
\node at (2.25,0.25) {\scriptsize $\mathbb{R}^{64}$};
\node at (4.2,0.25) {\scriptsize $\mathbb{R}^{64}$};
\node at (6.1,0.25) {\scriptsize $\mathbb{R}^{64}$};
\node at (8,0.25) {\scriptsize $\mathbb{R}^{64}$};
\node at (9.9,0.25) {\scriptsize $\mathbb{R}^{64}$};
\node at (11.8,0.25) {\scriptsize $\mathbb{R}^{64}$};
\node at (13.6,0.25) {\scriptsize $\mathbb{R}^{20r}$};
\end{tikzpicture}
\end{center}
Here, ``affine" refers to an affine layer performing the transformation ${\bf x}\rightarrow W{\bf x}+{\bf b}$ with weight $W$ and bias ${\bf b}$, ``SELU" is the standard scaled exponential linear unit activation function and ``softmax" is a softmax layer  which ensures that the output can be interpreted as a vector of probabilities which sum to one. The input of this network is the charge matrix $\mathcal{Q}$, in line with the input dimension of $10r$ while the output is a probability vector whose dimension, $20r$, equals the number of different actions~\eqref{action}.\\[2mm]
Training data is provided in batches which consist of triplets $(\mathcal{Q}_t,\alpha_t,G_t)$, where the actions $\alpha_t$ are represented by a standard unit vector in $\mathbb{R}^{20r}$. The probability of an action can then be written as $\pi_\theta(\mathcal{Q}_t,\alpha_t)=\alpha_t\cdot f_{\theta}(\mathcal{Q}_t)$ and the loss~\eqref{loss} takes the form
\begin{equation}
 L(\theta)= G_t\ln(\alpha_t\cdot f_{\theta}(\mathcal{Q}_t)\; .
\end{equation} 
Based on this loss function, the above network is trained with the ADAM optimiser, using batch sizes of $32$ and a typical learning rate of $\lambda=1/4000$.

\subsection{Agent}
The FN environment will be explored by a single agent, following episodes~\eqref{eppi} of maximal length $N_{\rm ep}=32$, and guided by the policy network $\pi_\theta$. Each new episode is started from a random state, to improve exploration of the environment.  Terminal states which are encountered during training are stored for later analysis.\\[2mm]
The FN environment and the REINFORCE algorithm are realised as MATHEMATICA~\cite{Mathematica} packages, the latter based on the MATHEMATICA  suite of machine learning modules. For terminal states found during training or by applying the trained network we perform a further Monte Carlo analysis in the space of order one coefficients $a_{ij}$, $b_{ij}$ (which were held fixed during training) in order to optimise their intrinsic value $\mathcal{V}(\mathcal{Q})$.
    
\section{Results}\label{sec:Training}
In this section, we present the results we have obtained by applying the REINFORCE algorithm to the FN environment, as described in the previous section. We focus on the two cases of one $U(1)$ symmetry with one singlet scalars and two $U(1)$ symmetries with two singlet scalars, starting with the former.

\subsection{One $U(1)$ symmetry}\label{sec:OneSymOneScalar}
 \begin{figure}
        \centering
        \subfloat[\sf Loss vs batch number.]{\includegraphics[width=0.43\linewidth]{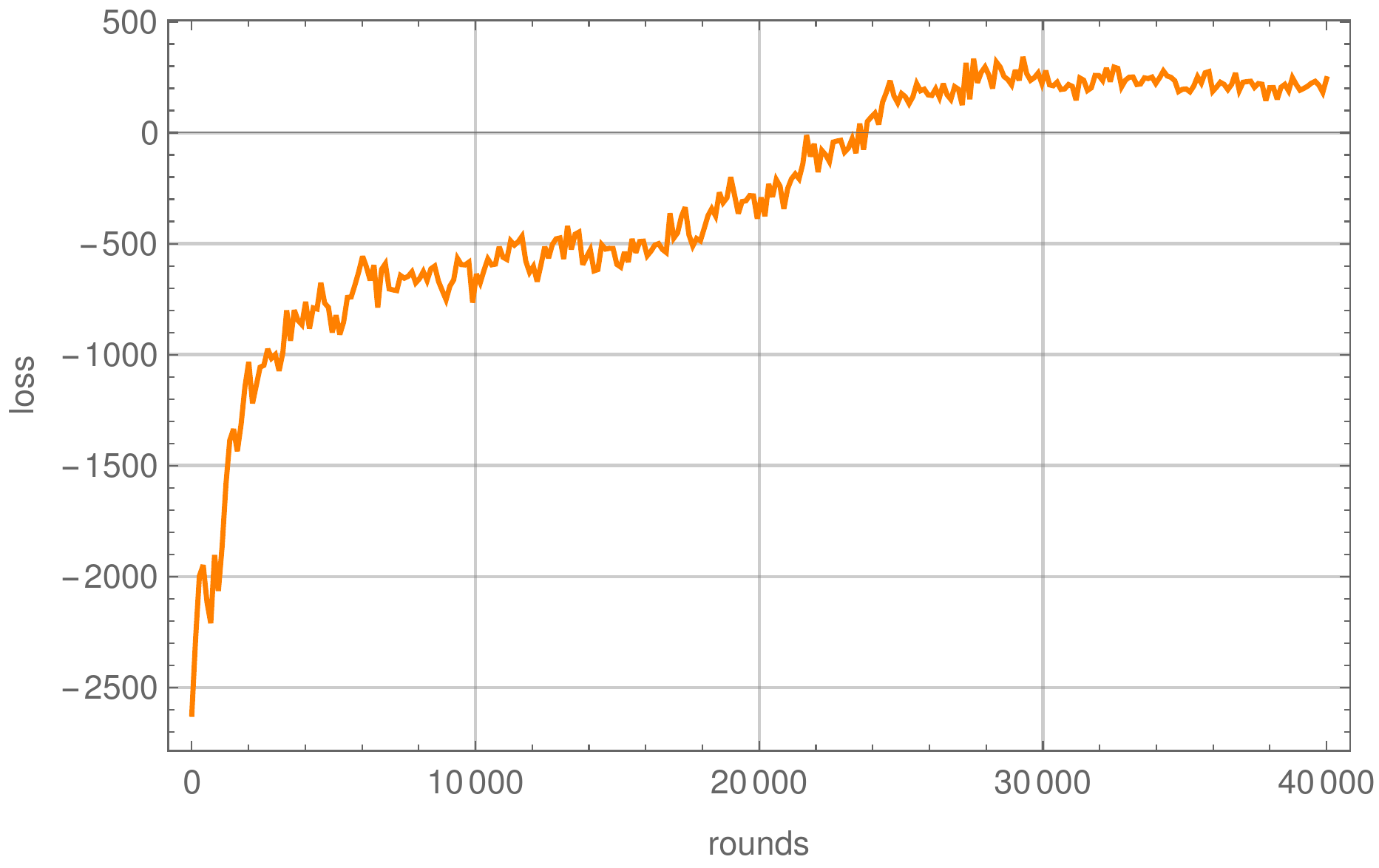}}\qquad
        \subfloat[\sf Return vs batch number.]{\includegraphics[width=0.43\linewidth]{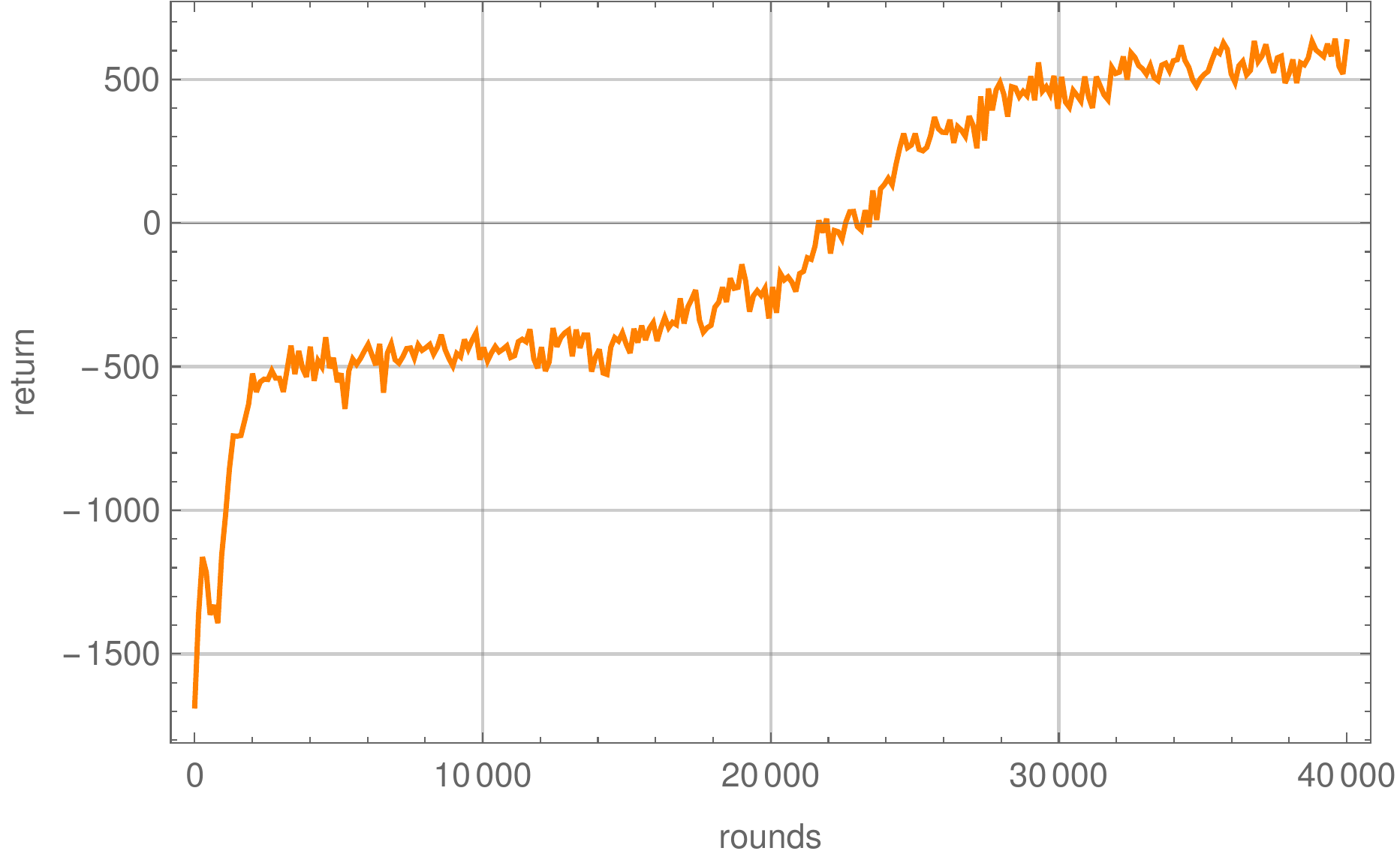}}\\
        \subfloat[\sf Fraction of terminal episodes vs episode number.]{\includegraphics[width=0.43\textwidth]{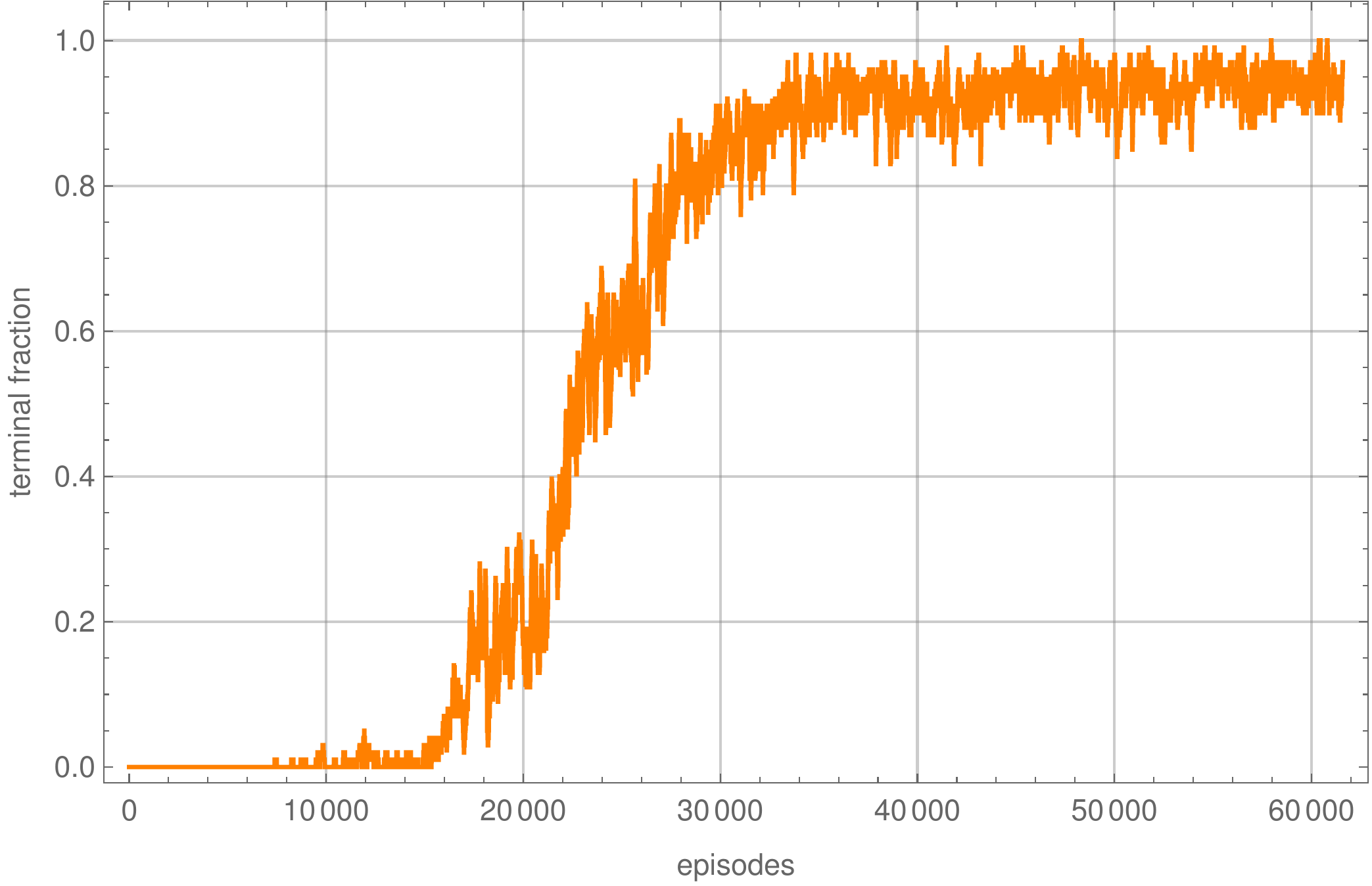}}\qquad%
        \subfloat[\sf Number of terminal states vs episode number.]{\includegraphics[width=0.43\textwidth]{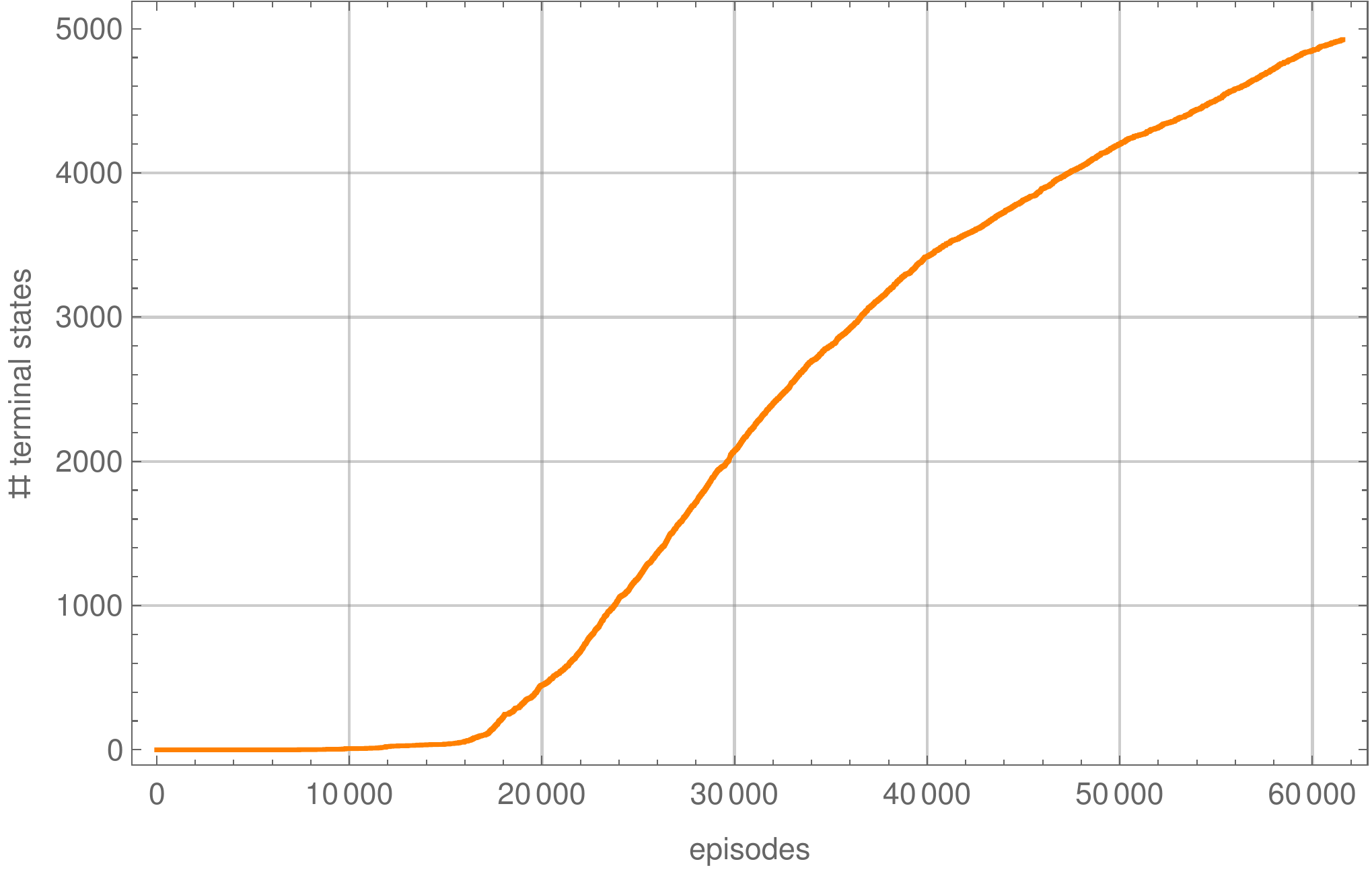}}%
        \caption{Training measurements for the case of a single $U(1)$ and $-q_{\rm min}=q_{\rm max}=9$.}
        \label{fig:1Scal1SymTrainingData}
 \end{figure}
\begin{figure}[!h]
        \centering
        \subfloat[\sf Intrinsic value (green) and reward (red) vs episode steps.]{\includegraphics[width=0.43\linewidth]{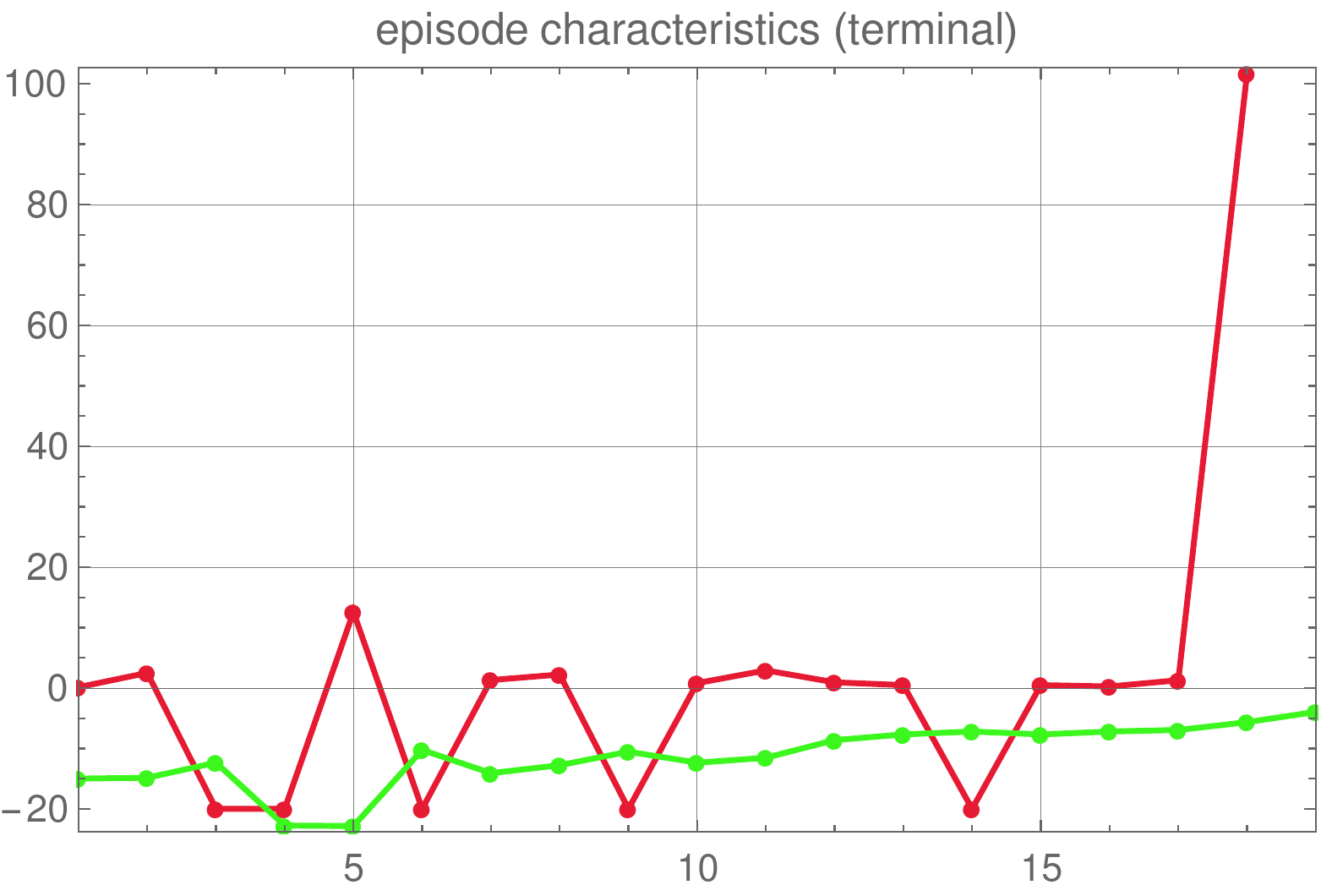}}\qquad
        \subfloat[\sf Two-dimensional projection of the charge matrices $\mathcal{Q}_t$ of the episode connecting the initial state~\eqref{initial1} (yellow dot) with the final state~\eqref{final1} (red dot). The labels indicate the intrinsic values.]{\includegraphics[width=0.43\linewidth]{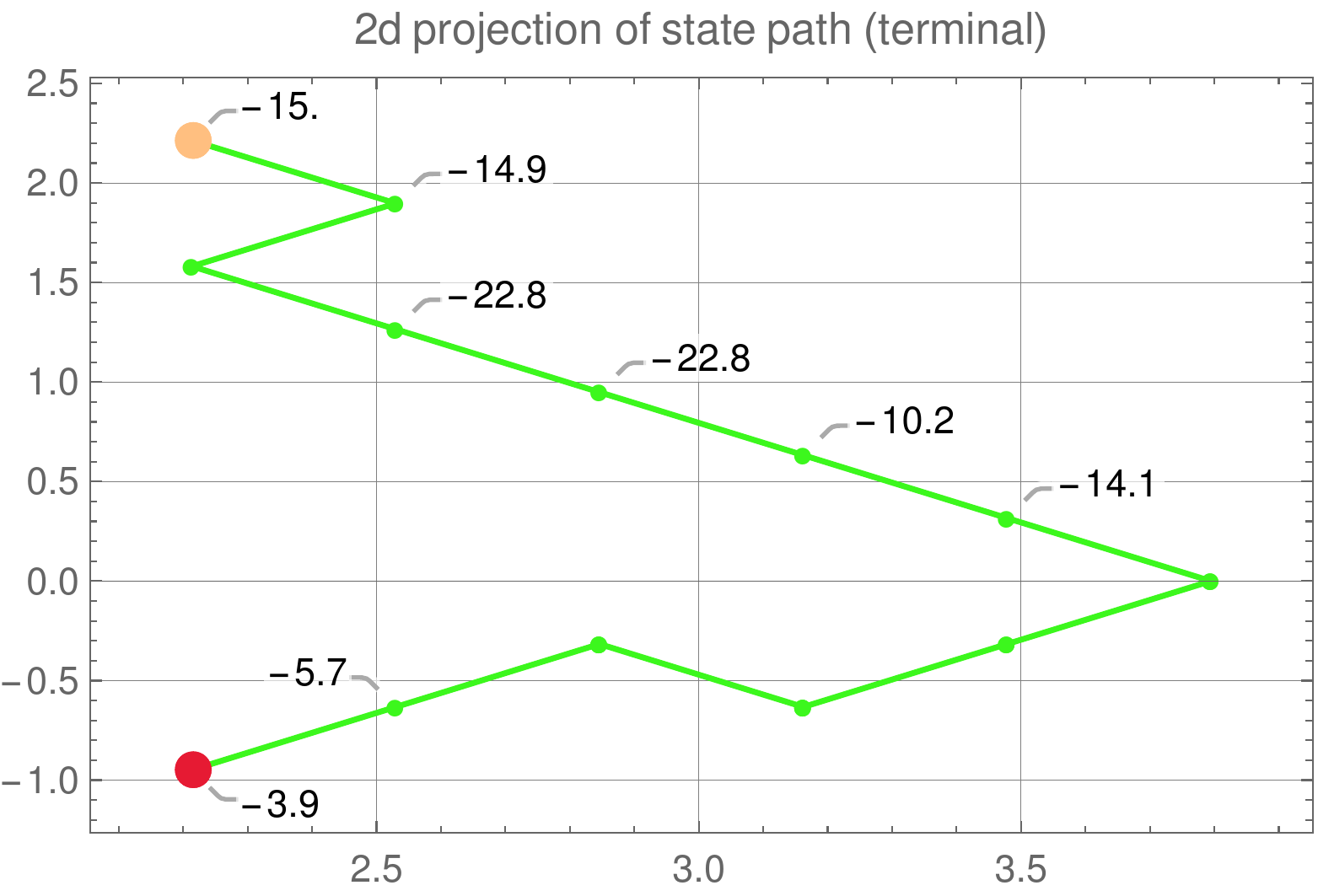}}
        \caption{Characteristics of the episode connecting the states~\eqref{initial1} and \eqref{final1}.}
        \label{fig:EpsiodePathExample}
\end{figure}
\begin{figure}[h]
        \centering
        \subfloat[\sf Intrinsic value (green) and reward (red) vs episode steps.]{\includegraphics[width=0.43\linewidth]{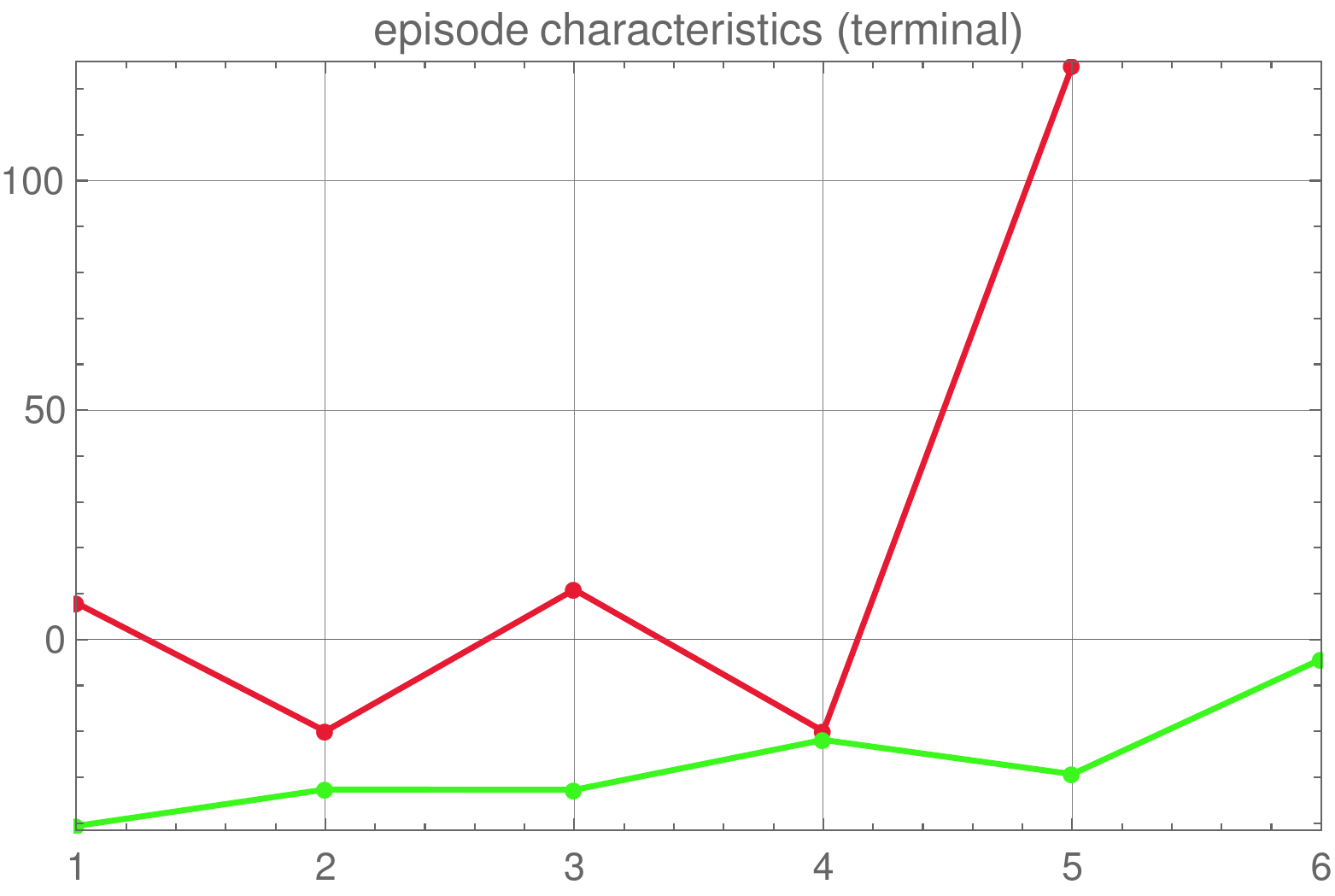}}\qquad
        \subfloat[\sf Two-dimensional projection of the charge matrices $\mathcal{Q}_t$ of the episode connecting the initial state~\eqref{initial2} (yellow dot) with the final state~\eqref{final2} (red dot). The labels indicate the intrinsic values.]{\includegraphics[width=0.43\linewidth]{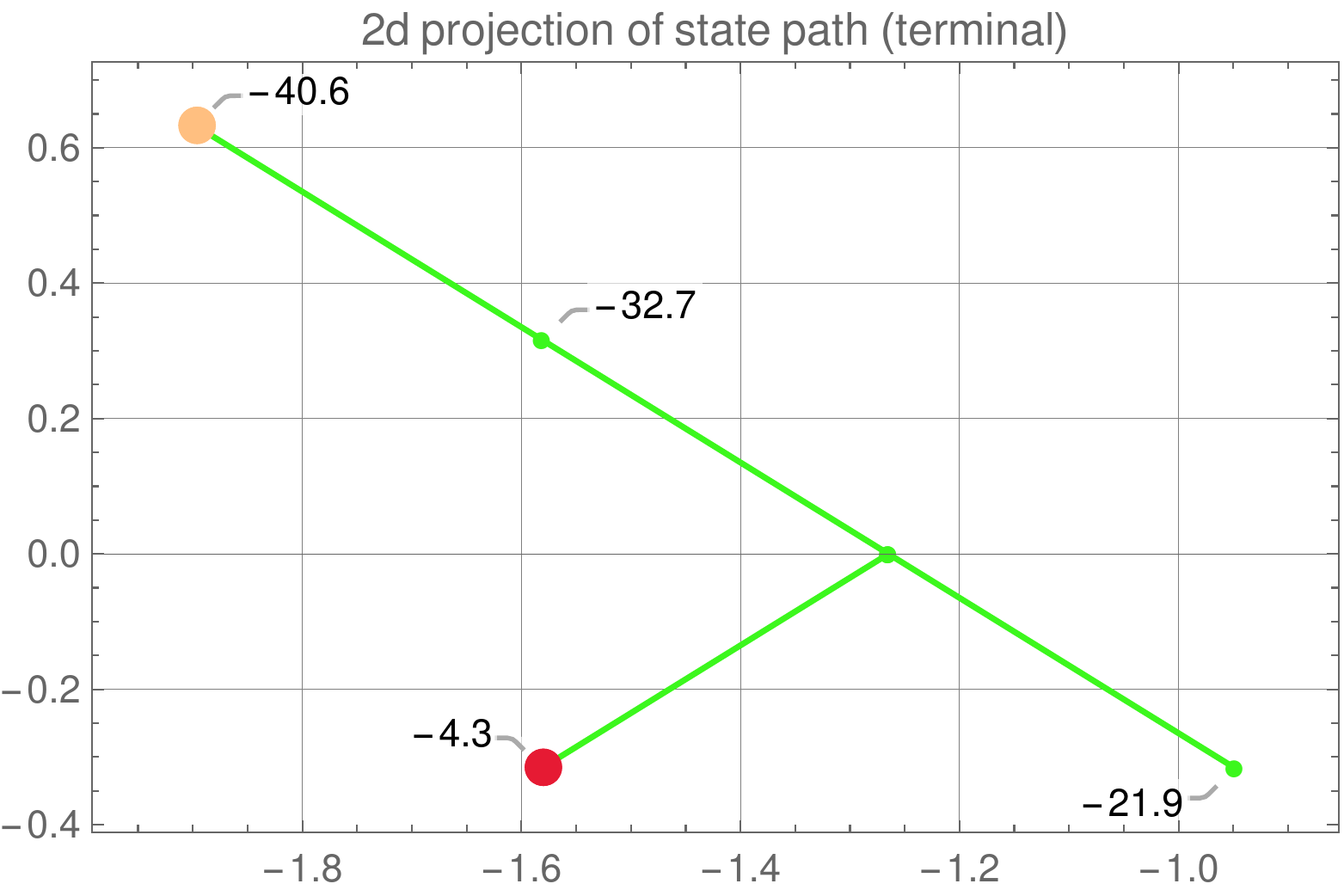}}
        \caption{Characteristics of the episode connecting the states~\eqref{initial2} and \eqref{final2}.}
        \label{fig:EpsiodePath1Scal1Sym}
 \end{figure}

 \noindent
The entries of the $1\times 10$ charge matrix $\mathcal{Q}$ are restricted as in Eq.~\eqref{envrange}, with $-q_{\rm min}=q_{\rm max}=9$, so the environment contains $19^{10}\sim 10^{13}$ states. Training of the network in Section~\ref{sec:nn} takes about an hour on a single CPU and the measurements made during training are shown in Fig.~\ref{fig:1Scal1SymTrainingData}. After an initial phase of exploration, lasting for about 15000 rounds, the network learns rapidly and the fraction of episodes which end in terminal states  (plot (c) in the Fig.~\ref{fig:1Scal1SymTrainingData}) rises to $>90\%$ within 10000 rounds or so. This pattern is quite characteristic and persists under variation of the various pieces of meta data, including the depth and width of the network, the constants which enter the definition~\eqref{reward} of the reward and the definition of a terminal state. The result is also stable under modest variations of the learning rate $\lambda=1/4000$, although too large learning rates ($\lambda>1/1000$) suppress exploration and lead to convergence to the ``wrong" policy. The residual positive loss in Fig.~\ref{fig:1Scal1SymTrainingData} (a) can be attributed to the fact that frequently more than one efficient path to a terminal state exists. In other words, there are several very similar optimal policies.\\[2mm]
During training, 4924 terminal states are found, which are reduced to 4630 after the redundancies due to the permutations~\eqref{psymm} are removed. Episodes guided by the trained network, starting at a random state and with maximal length $32$, lead to terminal states in $93\%$ of cases, and the average episode length is $16.4$.\\[2mm]
The intrinsic values of these $4630$ models found during training are optimised by performing a Monte-Carlo search over the order one coefficients $a_{ij}$, $b_{ij}$. In this way, we find $89$ models $\mathcal{Q}$ with an intrinsic value  $\mathcal{V}(\mathcal{Q})>-1$. From these, the model with the highest intrinsic value is given by~\footnote{Even though the Higgs charge is not part of the environment we include it here for convenience.}
\begin{equation}
 \mathcal{Q}=
      \left(
            \begin{array}{ccc|ccc|ccc|c|ccc}
                  Q_1 & Q_2 & Q_3 & u_1 & u_2 & u_3 & d_1 & d_2 & d_3 & H &  \phi  \\ \hline 
                 6 & 4 & 3 & -2 & 2 & 4 & -3 & -1 & -1 & 1 & 1
            \end{array}
        \right)\; .
\end{equation}
For a scalar VEV $v_1\simeq 0.224$ and the order one coefficients
\begin{equation}
    (a_{ij}) \simeq
        \left(
\begin{array}{rrr}
 -1.975 & 1.284 & -1.219 \\
 1.875 & -1.802 & -0.639 \\
 0.592 & 1.772 & 0.982 \\
\end{array}
\right)\;,\qquad
    (b_{ij}) \simeq 
       \left(
\begin{array}{rrr}
 -1.349 & 1.042 & 1.200 \\
 1.632 & 0.830 & -1.758 \\
 -1.259 & -1.085 & 1.949 \\
\end{array}
\right)\;,
\end{equation}
this model has an intrinsic value $\mathcal{V}(\mathcal{Q})\simeq -0.598$ and leads to the mass matrices
\begin{equation}
    M_{u} \simeq
        \left(
\begin{array}{rrr}
 0.000 & 0.126 & -2.380 \\
 0.009 & -3.517 & -24.904 \\
 0.013 & 15.456 & 170.815 \\
\end{array}
\right)\;,\qquad
    M_{d} \simeq 
        \left(
\begin{array}{rrr}
 -0.001 & 0.023 & 0.026 \\
 0.036 & 0.363 & -0.768 \\
 -0.123 & -2.119 & 3.806 \\
\end{array}
\right)\; .
\end{equation}
After diagonalisation, the resulting quark masses and mixings are
\begin{equation}
 \begin{array}{lll}
 (m_u,m_c,m_t)&\simeq& (0.003, 1.292, 173.358)\,{\rm GeV}\\
 (m_d,m_s,m_b)&\simeq& (0.005, 0.066, 4.439)\,{\rm GeV}
 \end{array}\;,\qquad
  V_{CKM} = \left(
\begin{array}{rrr}
 0.969 & 0.247 & 0.003 \\
 -0.247 & 0.968 & 0.050 \\
 0.009 & -0.049 & 0.999 \\
\end{array}
\right)\; ,
\end{equation} 
in reasonable agreement with the values in Table~\ref{tab:exp} and Eq.~\eqref{eqn:CKMVal}. Further examples of models with a high intrinsic value found during training are listed in Appendix~\ref{app:A}.\\[2mm]
Of course, the trained network can be used to find new models. For example, consider starting with the initial state
\begin{equation}\label{initial1}
\mathcal{Q}=
        \left(
            \begin{array}{ccc|ccc|ccc|c|ccc}
                Q_1 & Q_2 & Q_3 & u_1 & u_2 & u_3 & d_1 & d_2 & d_3 & H &  \phi  \\ \hline 
                0 & 2 & 0 & 0 & 4 & 0 & 0 & 0 & 0 & 0 & 1
            \end{array}
        \right)\; .
\end{equation} 
 The optimal intrinsic value for this state, achieved for a singlet VEV $v_1\simeq 0.112$, is $\mathcal{V}(\mathcal{Q})\simeq -15$, so this is definitely not a phenomenologically viable model. Using \eqref{initial1} as the initial state of an episode, guided by the trained network, it takes $18$ steps to reach the terminal state
\begin{equation}\label{final1}
\mathcal{Q}=
        \left(
            \begin{array}{ccc|ccc|ccc|c|ccc}
                 Q_1 & Q_2 & Q_3 & u_1 & u_2 & u_3 & d_1 & d_2 & d_3 & H &  \phi  \\ \hline 
                 2 & 3 & 1 & 1 & 3 & 3 & -2 & -2 & -3 & 2 & 1
            \end{array}
        \right)\; ,
\end{equation}
with intrinsic value $\mathcal{V}(\mathcal{Q})\simeq -3.94$ for a singlet VEV $v_1\simeq 0.056$. The intrinsic value and the reward along this episode, as well as a two-dimensional projection of the path mapped out by the episode is shown in Fig.~\ref{fig:EpsiodePathExample}.\\[2mm]
We can also test the trained network by checking whether it can guide us towards a model known in the literature, starting at a nearby state.
For example, consider the model from Ref.~\cite{MassMatrixTwo}, given by the charge matrix
\begin{equation}\label{final2}
  \mathcal{Q}=
        \left(
            \begin{array}{ccc|ccc|ccc|c|ccc}
                 Q_1 & Q_2 & Q_3 & u_1 & u_2 & u_3 & d_1 & d_2 & d_3 & H &  \phi  \\ \hline 
                3 & 2 & 0 & -3 & -1 & 0 & -3 & -2 & -2 & 0 & 1
            \end{array}
        \right)\; .
\end{equation}
which has an intrinsic value of $\mathcal{V}(\mathcal{Q})\simeq -4.3$ for a singlet VEV $v_1\simeq 0.159$. Suppose we use the initial state
\begin{equation}\label{initial2}
  \mathcal{Q}=
        \left(
            \begin{array}{ccc|ccc|ccc|c|ccc}
                 Q_1 & Q_2 & Q_3 & u_1 & u_2 & u_3 & d_1 & d_2 & d_3 & H &  \phi  \\ \hline 
                 4 & 2 & -3 & -3 & -1 & 0 & -3 & -2 & -2 & -3 & 2
            \end{array}
        \right)\; .
\end{equation}
which is a perturbation of the literature model~\eqref{final2} but, as is, does not amount to a potentially viable model. Generating an episode starting at the state~\eqref{initial2} then leads to the literature model~\eqref{final2} in four steps, as indicated in Fig.~\ref{fig:EpsiodePath1Scal1Sym}.

\subsection{Two $U(1)$ symmetries}\label{sec:TwoSymTwoScalar}
\begin{figure}[h]
        \centering
        \subfloat[\sf Loss vs batch number.]{\includegraphics[width=0.43\linewidth]{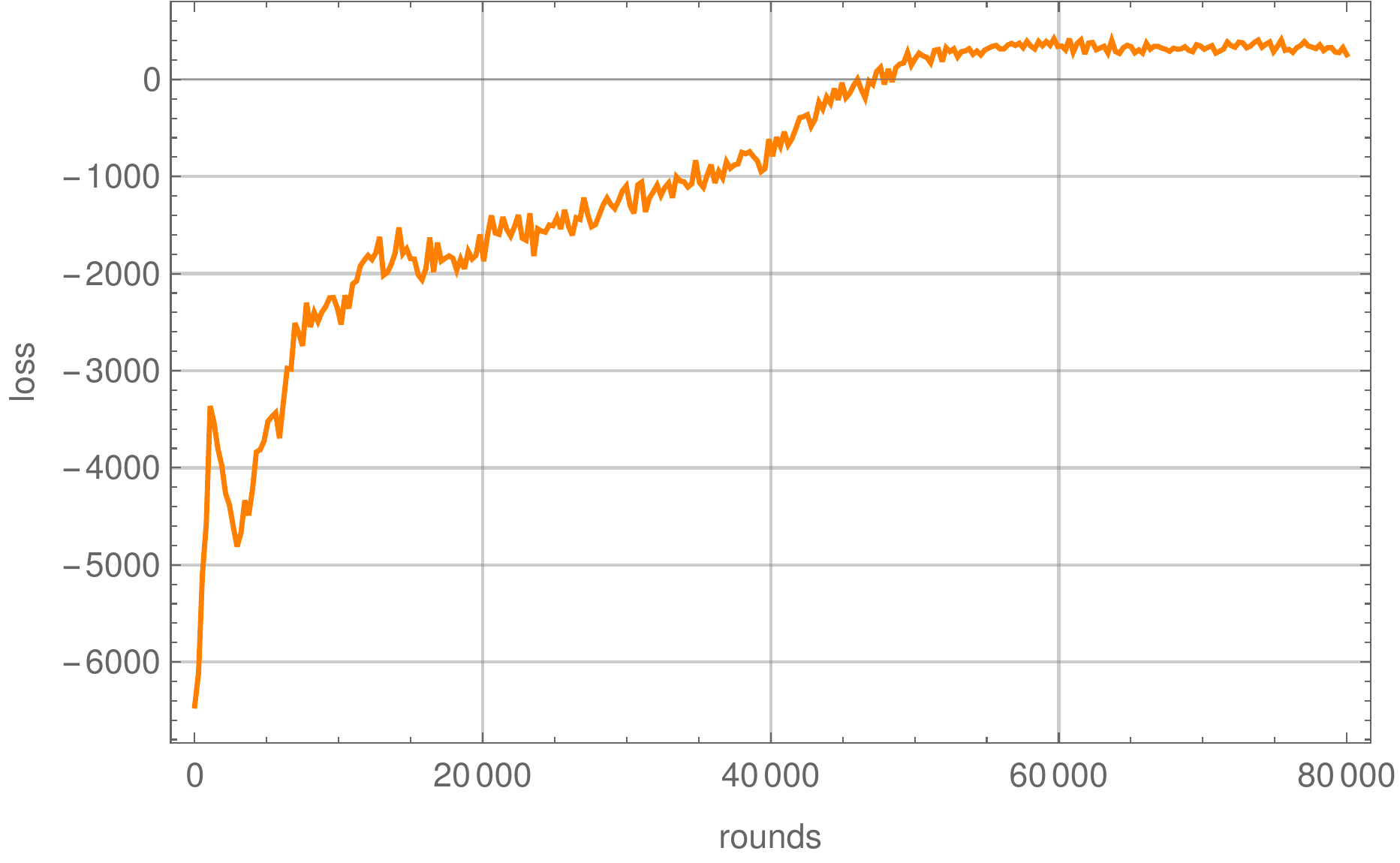}}\qquad
        \subfloat[\sf Return vs batch number.]{\includegraphics[width=0.43\linewidth]{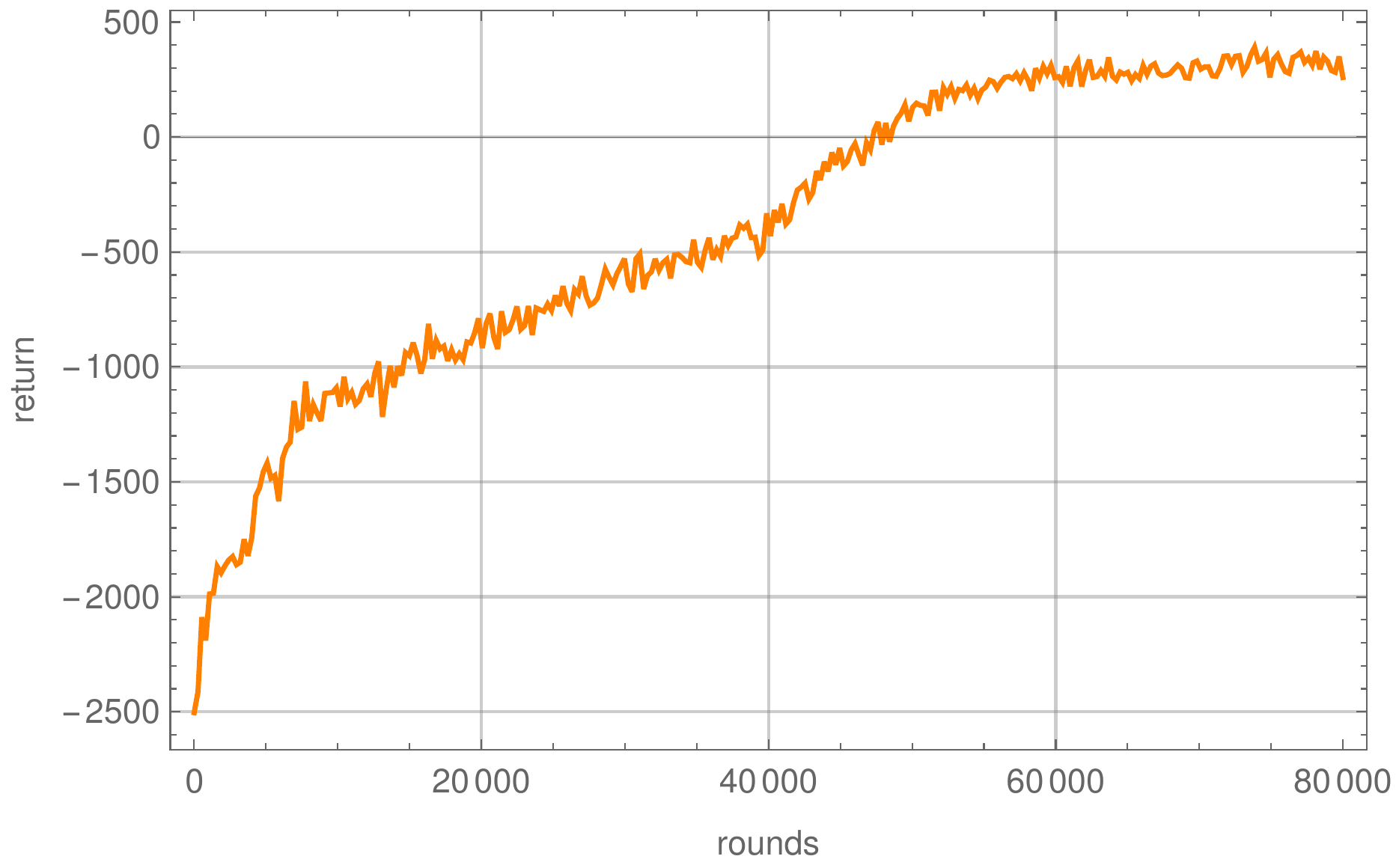}}\\
        \subfloat[\sf Fractional of terminal episodes vs episode number.]{\includegraphics[width=0.43\textwidth]{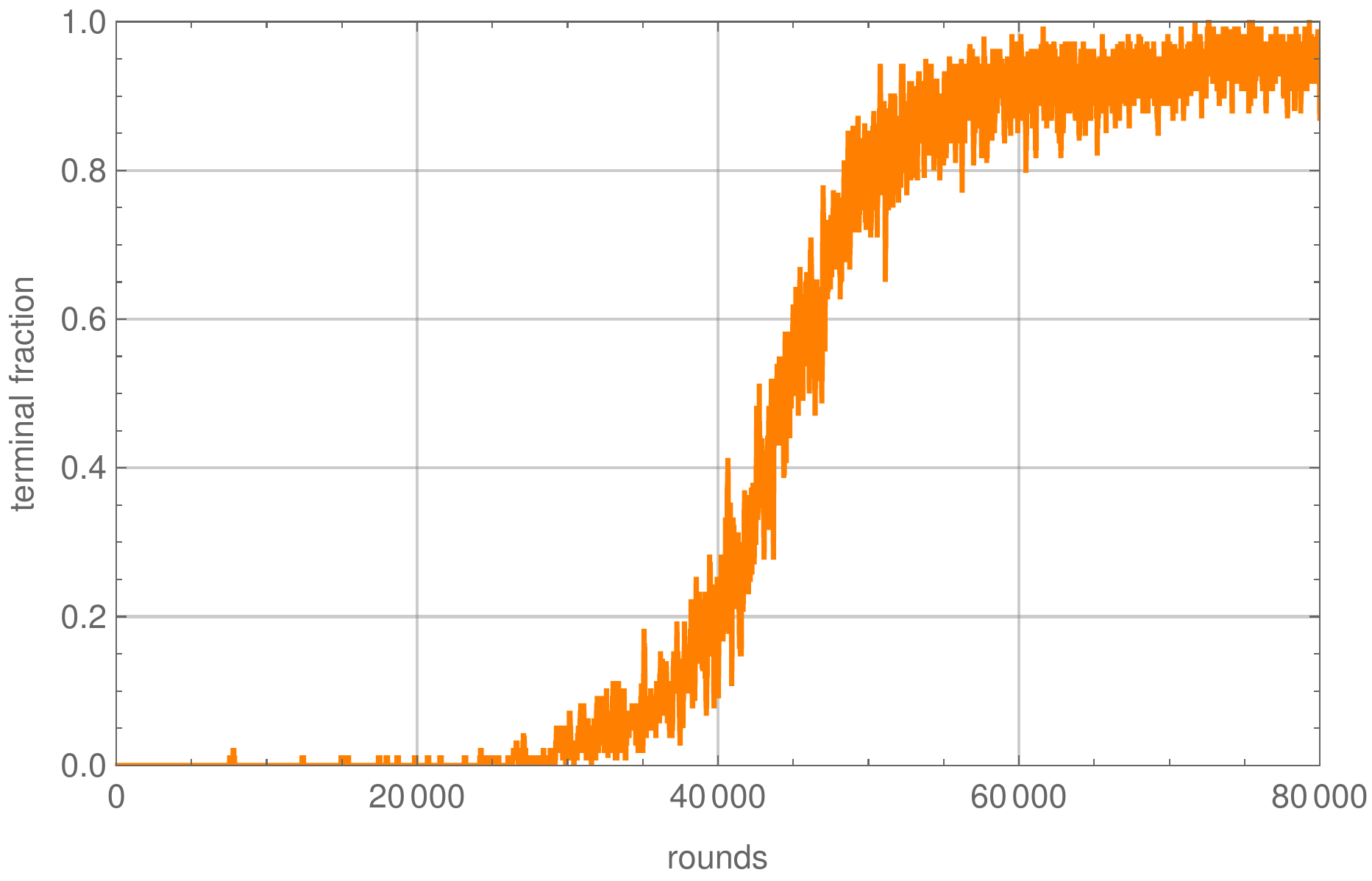}}\qquad%
        \subfloat[\sf Number of terminal states vs episode number.]{\includegraphics[width=0.43\textwidth]{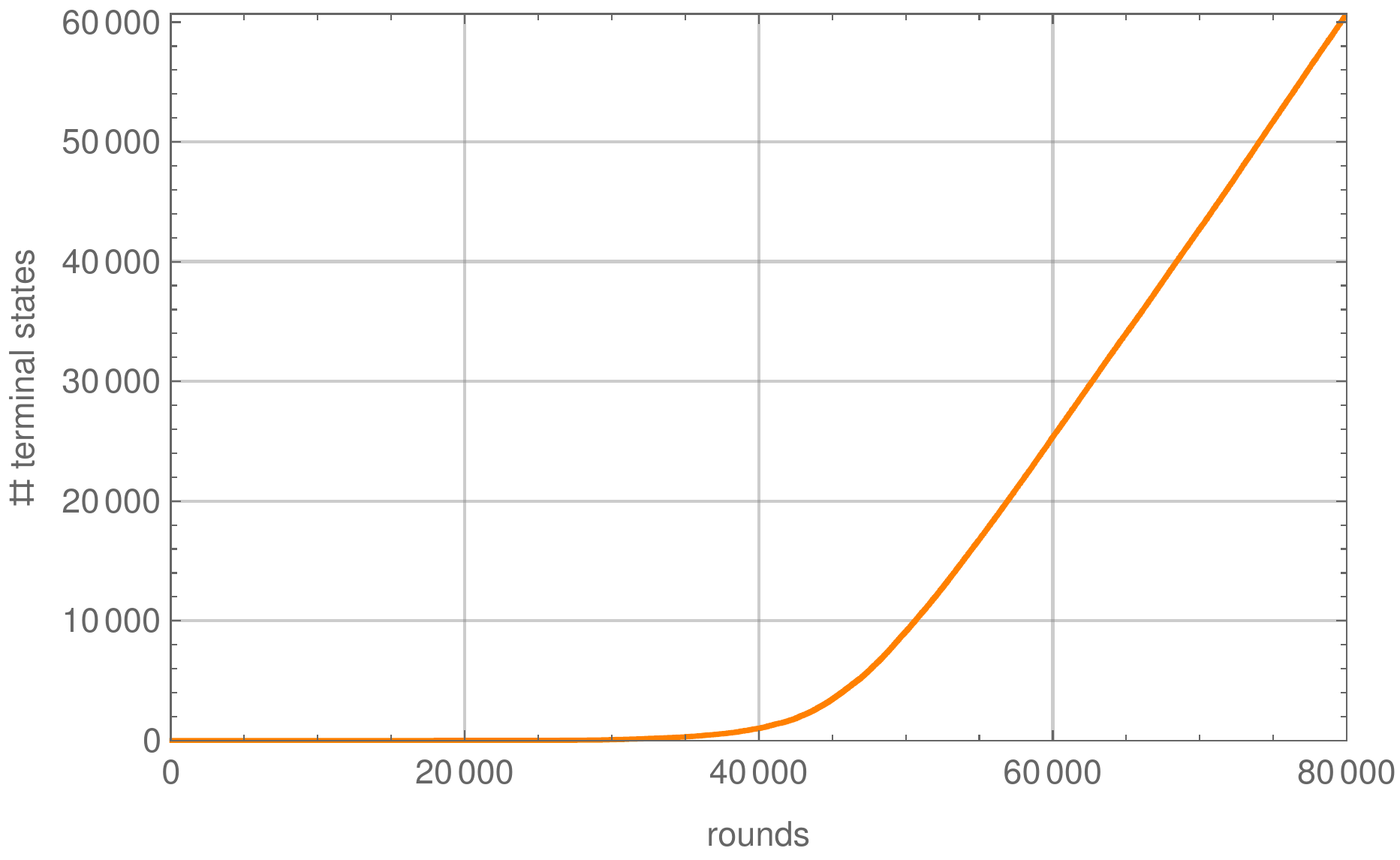}}%
        \caption{Training measurements for the case of a two $U(1)$ symmetries and $-q_{\rm min}=q_{\rm max}=5$.}
        \label{fig:2Scal2SymTrainingData}
 \end{figure}

\begin{figure}
        \centering
        \subfloat[[\sf Intrinsic value (green) and reward (red) vs episode steps.]{\includegraphics[width=0.43\linewidth]{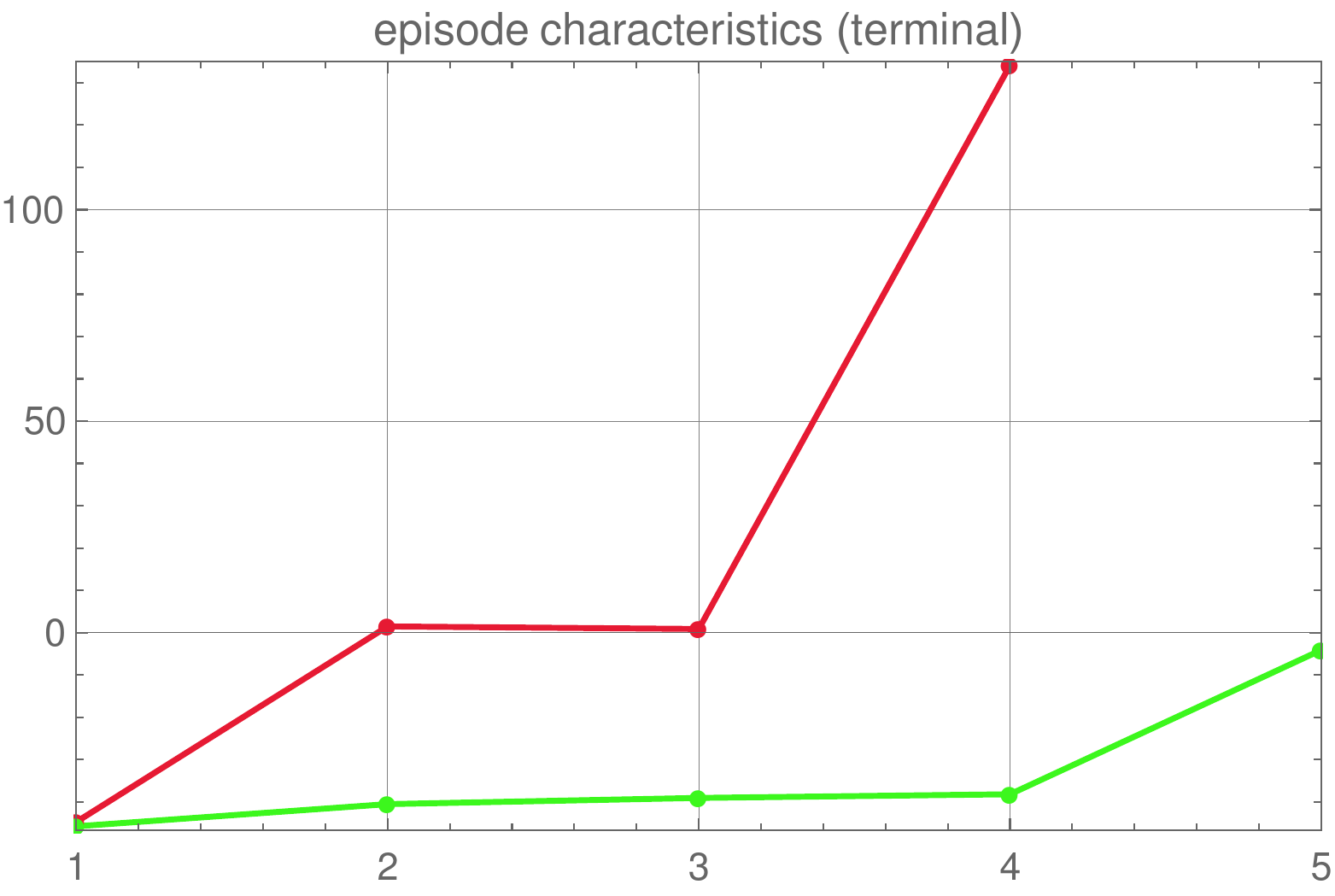}}\qquad
        \subfloat[\sf Two-dimensional projection of the charge matrices $\mathcal{Q}_t$ of the episode connecting the initial state~\eqref{initial3} (yellow dot) with the final state~\eqref{final3} (red dot). The labels indicate the intrinsic values.]{\includegraphics[width=0.43\linewidth]{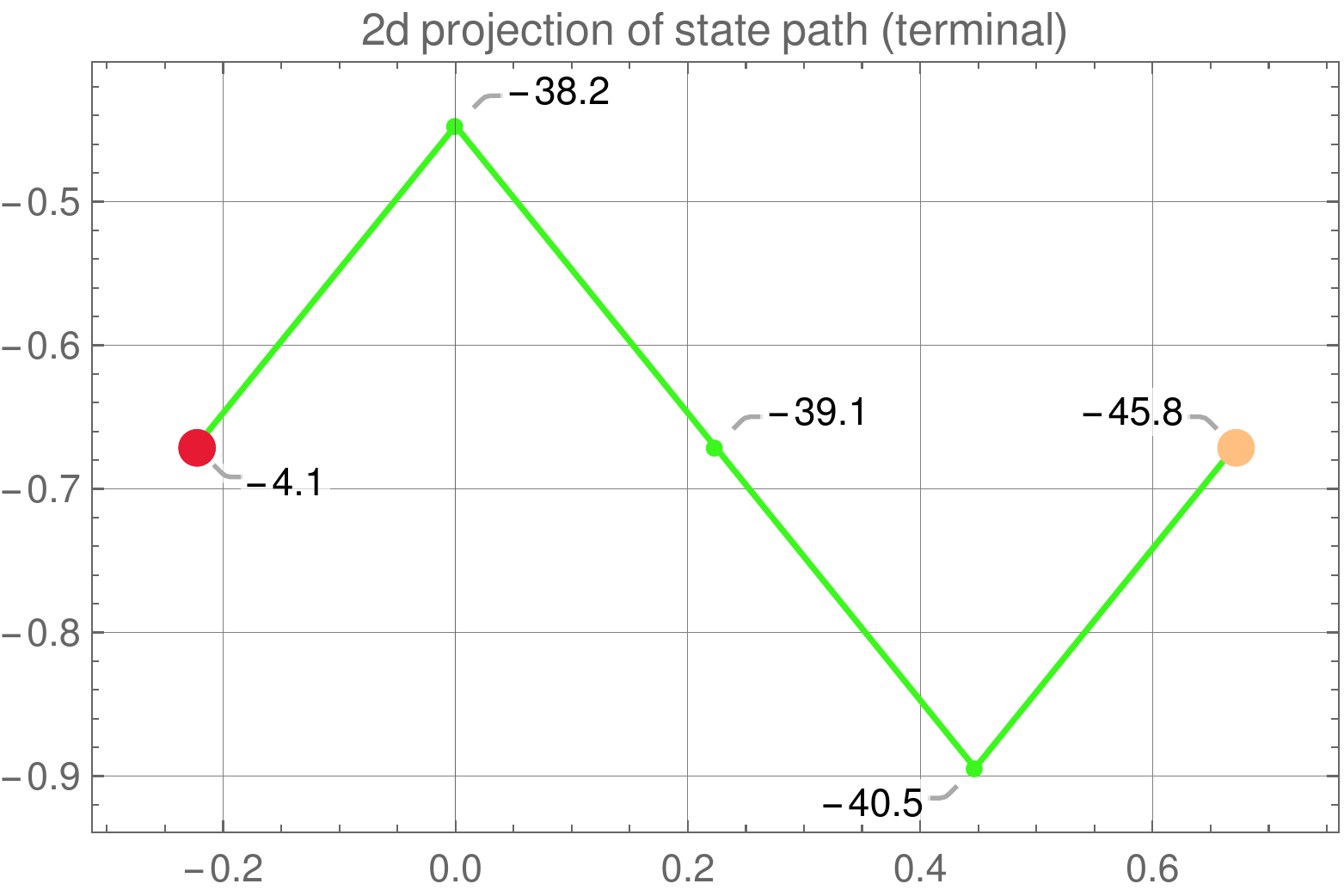}}
        \caption{Characteristics of the episode connecting the states~\eqref{initial3} and \eqref{final3}.}
        \label{fig:EpsiodePath2Scal2Sym}
\end{figure}

Next, we present results for an environment with two $U(1)$ symmetries and two singlet scalar fields. The entries of the $2\times 10$ charge matrix $\mathcal{Q}$ are constrained as in Eq.~\eqref{envrange} but we now consider a somewhat smaller range with $-q_{\rm min}=q_{\rm max}=5$. This still leads to a considerably larger environment than previously, with a total of $11^{20}\sim 10^{21}$ states.\\[2mm]
Training for this environment on a single CPU takes about $25$ hours and leads to the measurements shown in Fig.~\ref{fig:2Scal2SymTrainingData}. The networks finds $60686$ terminal states which reduce to $57807$ once the permutation redundancies~\eqref{psymm} are removed. Episodes guided  by the trained network and with maximal length $32$  lead to terminal states in 95\% of cases and the average episode length is 19.9 steps.\\[2mm]
After a Monte-Carlo optimisation of the order one coefficients $a_{ij}$, $b_{ij}$ we find $2019$ from the $57807$ models found during training have an intrinsic value $\mathcal{V}(\mathcal{Q})>-1$.  The best of these has charge allocation
\begin{equation}
    \mathcal{Q}=
        \left(
            \begin{array}{rrr|rrr|rrr|r|rrr}
                 Q_1 & Q_2 & Q_3 & u_1 & u_2 & u_3 & d_1 & d_2 & d_3 & H &  \phi_1 & \phi_2 \\ \hline 
                 2 & 2 & 1 & -2 & 0 & 1 & -1 & 0 & 1 & 0 & 1 & 0\\  
                 1 & 0 & 0 & 0 & 0 & 0 & -1 & -1 & -2 & 0 & 0 & 1
            \end{array}
        \right)\; ,
    \end{equation}
and an intrinsic value $\mathcal{V}(\mathcal{Q})\simeq -0.390$, provided we choose singlet VEVs $v_1\simeq 0.079$, $v_2\simeq 0.112$ and order one coefficients
\begin{equation}
    (a_{ij}) \simeq
        \left(
\begin{array}{rrr}
 -1.898 & 0.834 & -0.587 \\
 -0.575 & -0.592 & 1.324 \\
 -1.123 & -1.265 & 0.982 \\
\end{array}
\right)\;,\qquad
    (b_{ij}) \simeq 
        \left(
\begin{array}{rrr}
 -1.759 & 1.358 & 1.013 \\
 -1.267 & 1.897 & -1.196 \\
 1.771 & 1.386 & -1.785 \\
\end{array}
\right)\; .
\end{equation}
This results in the mass matrices
\begin{equation}
    M_{u}\simeq
        \left(
\begin{array}{rrr}
 -0.001 & 0.103 & -0.910 \\
 -0.004 & -0.650 & 18.297 \\
 -0.098 & -17.489 & 170.815 \\
\end{array}
\right)\;,\qquad
    M_{d} \simeq 
        \left(
\begin{array}{rrr}
 -0.002 & 0.019 & 0.020 \\
 -0.012 & 0.234 & -0.208 \\
 0.218 & 2.149 & -3.910 \\
\end{array}
\right)\; ,
\end{equation}
and the masses and mixing
\begin{equation}
 \begin{array}{lll}
    (m_u,m_c,m_t)&\simeq& (0.002,1.210,172.679)\,{\rm GeV}\\
    (m_d,m_s,m_b)&\simeq& (0.005,0.111,4.476)\,{\rm GeV}
  \end{array}\;,\qquad
  V_{\rm CKM} = \left(
\begin{array}{rrr}
 0.975 & -0.223 & 0.004 \\
 0.223 & 0.974 & 0.040 \\
 -0.013 & -0.038 & 0.999 \\
\end{array}
\right)\; ,
\end{equation}
in rough agreement with the values in Table~\ref{eqn:CKMVal}. More examples of promising models found by the network are listed in Appendix~\ref{app:B}.\\[2mm]
We can also demonstrate that the trained network is capable of finding models which have been constructed in the literature. Consider the model from Ref.~\cite{MassMatrixTwo} which is described by the charge matrix
\begin{equation}\label{final3}
  \mathcal{Q}=
        \left(
            \begin{array}{rrr|rrr|rrr|r|rrr}
                 Q_1 & Q_2 & Q_3 & u_1 & u_2 & u_3 & d_1 & d_2 & d_3 & H &  \phi_1 & \phi_2 \\ \hline 
                 3 & 0 & 0 & 1 & -1 & 0 & 1 & -4 & 0 & 0 & 1 & 0\\  
                 0 & 1 & 0 & -2 & 0 & 0 & -2 & 1 & -1 & 0 & 0 & 1
            \end{array}
        \right)\; .
\end{equation}
For singlet VEVs $v_1\simeq 0.158$ and $v_2\simeq 0.028$ it is a terminal state with intrinsic value $\mathcal{V}(\mathcal{Q})\simeq -4.1$ which, however, has not been found during training. To see that this model can be obtained we start an episode at a nearby state with charge matrix
\begin{equation}\label{initial3}
  \mathcal{Q}=
        \left(
            \begin{array}{rrr|rrr|rrr|r|rrr}
                Q_1 & Q_2 & Q_3 & u_1 & u_2 & u_3 & d_1 & d_2 & d_3 & H &  \phi_1 & \phi_2 \\ \hline 
                5 & 0 & 0 & 1 & -1 & 0 & 1 & -4 & 0 & 0 & 1 & 0\\  
                0 & 1 & 0 & -2 & 0 & 0 & -2 & 1 & -1 & 0 & 0 & 3
            \end{array}
        \right)\; .
\end{equation}
The trained network then takes us from this state to the literature model~\eqref{final3} in three steps, as can be seen in Fig.~\ref{fig:EpsiodePath2Scal2Sym}.

\section{Conclusion and outlook}
In this paper, we have studied particle physics model building with reinforcement learning (RL). We have focused on a simple model building framework - Froggatt-Nielsen (FN) models for quark masses and mixing - and the simplest policy-based RL algorithm. Our results show that successful model-building is indeed possible in this way. For both cases we consider, that is, for FN models with one $U(1)$ symmetries and two $U(1)$ symmetries, the network can be trained to settle on a highly efficient policy which leads to terminal states in $> 90\%$ of all cases and in an average number of $<20$ steps. The trained networks can be used to find promising models from random initial states and it is capable of finding literature models, provided it is started at a near-by state.\\[2mm]
There are numerous extensions of this work. At the most basic level, the order one coefficients which enter FN model building might be incorporated into the RL environment, rather than being fixed to random values as we have done here for simplicity. It would also be desirable to study the performance of other RL algorithms, such as actor-critic set-ups and Q-learning, on the FN environment. Extending the environment to include lepton masses is another interesting direction. More generally, we can ask if other areas of particle physics model building can be approached in this way. For example, can RL be used for dark matter model building?\\[2mm]
As its most ambitious, this line of thought suggests an RL environment which consists of large classes of quantum field theories, for example a large class of extensions of the standard model of particle physics, prescribed in some appropriate manner. The intrinsic value of such models might be determined by comparing their predictions with a wide range of experimental data. Realising such an environment would require significantly more theoretical preparation than was necessary for the FN environment as well as more computing power. However, the benefits of such a system might be considerable - it would allow exploring large classes of standard model extensions and their consistency with experimental data and might help to find the correct path for physics beyond the standard model.
    
\section*{Acknowledgments}
T.~R.~H.~is supported by an STFC studentship. A.~L.~would like to thank Andrei Constantin for useful discussions.
    
\appendix
\section{Example models for one $U(1)$ symmetry}\label{app:A}
 In this appendix we list some models with a single $U(1$) symmetry with a high intrinsic value $\mathcal{V}(\mathcal{Q})$, found during training.   
 \begin{longtable}{|c|c|} \hline
             charges & 
                $\mathcal{Q}=\left(
                    \begin{array}{c|ccc|ccc|ccc|c|ccc}
                         {} & Q_1 & Q_2 & Q_3 & u_1 & u_2 & u_3 & d_1 & d_2 & d_3 & H &  \phi  \\ \hline 
                        q & 6 & 4 & 3 & -2 & 2 & 4 & -3 & -1 & -1 & 1 & 1
                    \end{array}
                \right)$
            
            \\ \hline
             $\mathcal{O}(1)$ coeff. & 
                $(a_{ij}) \simeq
                \left(
\begin{array}{rrr}
 -1.975 & 1.284 & -1.219 \\
 1.875 & -1.802 & -0.639 \\
 0.592 & 1.772 & 0.982 \\
\end{array}
\right)\quad
                (b_{ij}) \simeq 
                \left(
\begin{array}{rrr}
 -1.349 & 1.042 & 1.200 \\
 1.632 & 0.830 & -1.758 \\
 -1.259 & -1.085 & 1.949 \\
\end{array}
\right)$
            \\ \hline
            VEV, Value &$v_1\simeq 0.224\;,\qquad \mathcal{V}(\mathcal{Q})\simeq -0.598$\\ \hline\hline
            charges & 
                $\mathcal{Q}=\left(
                    \begin{array}{ccc|ccc|ccc|c|ccc}
                         Q_1 & Q_2 & Q_3 & u_1 & u_2 & u_3 & d_1 & d_2 & d_3 & H &  \phi  \\ \hline 
                         1 & 2 & 0 & -1 & -3 & 1 & -3 & -5 & -4 & 1 & 1
                    \end{array}
                \right)$\\\hline
             $\mathcal{O}(1)$ coeff. & 
                $(a_{ij}) \simeq
                \left(
\begin{array}{rrr}
 -0.601 & 1.996 & 0.537 \\
 -0.976 & -1.498 & -1.156 \\
 1.513 & 1.565 & 0.982 \\
\end{array}
\right)\quad
                (b_{ij}) \simeq 
                \left(
\begin{array}{rrr}
 0.740 & -1.581 & -1.664 \\
 -1.199 & -1.383 & 0.542 \\
 0.968 & 0.679 & -1.153 \\
\end{array}
\right)$
            \\ \hline
            VEV, value & $v_1\simeq 0.158\;,\qquad  \mathcal{V}(\mathcal{Q})\simeq -0.621$\\ \hline\hline   
            charges & 
                $\mathcal{Q}=\left(
                    \begin{array}{ccc|ccc|ccc|c|ccc}
                         Q_1 & Q_2 & Q_3 & u_1 & u_2 & u_3 & d_1 & d_2 & d_3 & H &  \phi  \\ \hline 
                         7 & 8 & 4 & 2 & -1 & 4 & 0 & -1 & 0 & 0 & 1
                    \end{array}
                \right)$\\\hline
             $\mathcal{O}(1)$ coeff. & 
                $(a_{ij}) \simeq
                \left(
\begin{array}{rrr}
 -1.027 & -1.234 & 1.914 \\
 1.027 & -0.525 & -0.921 \\
 -1.995 & 1.317 & 0.982 \\
\end{array}
\right)\quad
                 (b_{ij})\simeq 
                \left(
\begin{array}{rrr}
 -1.223 & -1.322 & 1.121 \\
 1.004 & 0.945 & 1.851 \\
 -1.930 & -1.071 & -1.720 \\
\end{array}
\right)$
            \\ \hline
            VEV, value & $v_1\simeq 0.316\;,\qquad  \mathcal{V}(\mathcal{Q})\simeq -0.642$\\ \hline\hline
            charges & 
                $\mathcal{Q}=\left(
                    \begin{array}{ccc|ccc|ccc|c|ccc}
                          Q_1 & Q_2 & Q_3 & u_1 & u_2 & u_3 & d_1 & d_2 & d_3 & H &  \phi  \\ \hline 
                          3 & 4 & 1 & -3 & -6 & -1 & 1 & -1 & -1 & -2 & 1 
                    \end{array}
                \right)$\\\hline
             $\mathcal{O}(1)$ coefficients & 
                $(a_{ij})\simeq
                \left(
\begin{array}{rrr}
 1.226 & -0.747 & 1.017 \\
 1.473 & 1.350 & 1.776 \\
 -1.575 & 0.988 & 0.982 \\
\end{array}
\right)\quad
                (b_{ij}) \simeq 
                \left(
\begin{array}{rrr}
 -1.012 & 1.947 & -1.941 \\
 -1.834 & -1.410 & 1.170 \\
 -0.614 & -0.872 & 1.405 \\
\end{array}
\right)$
            \\ \hline
            VEV, value & $v_1\simeq 0.224\;,\qquad  \mathcal{V}(\mathcal{Q})\simeq -0.721$\\ \hline
\end{longtable}

\section{Example models for two $U(1)$ symmetries}\label{app:B}
In this appendix we list some models with a single $U(1$) symmetry with a high intrinsic value $\mathcal{V}(\mathcal{Q})$, found during training.  
\begin{longtable}{|c|c|} \hline
            charges & 
                $\mathcal{Q}=\left(
                    \begin{array}{rrr|rrr|rrr|r|rrr}
                         Q_1 & Q_2 & Q_3 & u_1 & u_2 & u_3 & d_1 & d_2 & d_3 & H &  \phi_1 & \phi_2  \\ \hline 
                         2 & 2 & 1 & -2 & 0 & 1 & -1 & 0 & 1 & 0 & 1 & 0 \\
                         1 & 0 & 0 & 0 & 0 & 0 & -1 & -1 & -2 & 0 & 0 & 1
                    \end{array}
                \right)$
            
            \\ \hline
              $\mathcal{O}(1)$ coeff.& 
                $(a_{ij})\simeq
                \left(
\begin{array}{rrr}
 -1.898 & 0.834 & -0.587 \\
 -0.575 & -0.592 & 1.324 \\
 -1.123 & -1.265 & 0.982 \\
\end{array}
\right)\quad
                (b_{ij})\simeq 
                \left(
\begin{array}{rrr}
 -1.759 & 1.358 & 1.013 \\
 -1.267 & 1.897 & -1.196 \\
 1.771 & 1.386 & -1.785 \\
\end{array}
\right)$
            \\ \hline
            VEVs, value &$(v_1,v_2)\simeq (0.079,0.112)\;,\qquad\mathcal{V}(\mathcal{Q})\simeq -0.390$\\ \hline\hline
             charges & 
                $\mathcal{Q}=\left(
                    \begin{array}{rrr|rrr|rrr|r|rrr}
                        Q_1 & Q_2 & Q_3 & u_1 & u_2 & u_3 & d_1 & d_2 & d_3 & H &  \phi_1 & \phi_2  \\ \hline 
                        3 & 3 & 0 & 1 & 0 & 1 & -2 & -1 & -1 & 1 & 1 & 0 \\
                        1 & 2 & 2 & 1 & -2 & 2 & 0 & -2 & -1 & 0 & 0 & 1
                    \end{array}
                \right)$
            
            \\ \hline
             $\mathcal{O}(1)$ coeff. & 
                $(a_{ij})\simeq
                \left(
\begin{array}{rrr}
 0.715 & -1.366 & -1.988 \\
 1.005 & 1.497 & 0.576 \\
 1.767 & -1.194 & 0.982 \\
\end{array}
\right)\quad
                (b_{ij})\simeq 
                \left(
\begin{array}{rrr}
 1.195 & -1.352 & -1.410 \\
 -0.979 & -1.860 & -0.521 \\
 -1.932 & -0.946 & 0.912 \\
\end{array}
\right)$
            \\ \hline
            VEVs, value &$(v_1,v_2)\simeq (0.224,0.224)\;,\qquad \mathcal{V}(\mathcal{Q})\simeq -0.425$\\ \hline\hline
            charges & 
                $\mathbb{Q}=\left(
                    \begin{array}{rrr|rrr|rrr|r|rrr}
                        Q_1 & Q_2 & Q_3 & u_1 & u_2 & u_3 & d_1 & d_2 & d_3 & H &  \phi_1 & \phi_2  \\ \hline 
                        2 & 3 & 1 & 1 & -1 & 1 & -1 & -1 & -2 & 0 & 1 & 0 \\
                        2 & 2 & 1 & -1 & -2 & 1 & 0 & -1 & 0 & 0 & 0 & 1
                    \end{array}
                \right)$
            
            \\ \hline
              $\mathcal{O}(1)$ coeff. & 
                $(a)_{ij})\simeq
                \left(
\begin{array}{rrr}
 1.603 & 1.793 & 1.589 \\
 0.653 & -1.887 & -0.869 \\
 -1.609 & -0.679 & 0.982 \\
\end{array}
\right)\quad
                (b_{ij})\simeq 
                \left(
\begin{array}{rrr}
 1.369 & -1.107 & 1.151 \\
 -1.658 & -1.831 & 1.868 \\
 1.819 & 1.395 & -0.813 \\
\end{array}
\right)$
            \\ \hline
            VEVs, Value &$(v_1,v_2)\simeq (0.316,0.158)\;,\qquad \mathcal{V}(\mathcal{Q})\simeq -0.480$\\ \hline\hline
            charges & 
                $\mathcal{Q}=\left(
                    \begin{array}{rrr|rrr|rrr|r|rrr}
                        Q_1 & Q_2 & Q_3 & u_1 & u_2 & u_3 & d_1 & d_2 & d_3 & H &  \phi_1 & \phi_2  \\ \hline 
                        1 & 3 & 1 & 0 & 0 & 0 & 0 & 1 & 1 & -1 & 1 & 0 \\
                        3 & 1 & 1 & -1 & 0 & 1 & 0 & 0 & 0 & 0 & 0 & 1
                    \end{array}
                \right)$
            
            \\ \hline
              $\mathcal{O}(1)$ coeff. & 
                $(a_{ij})\simeq
                \left(
\begin{array}{rrr}
 -1.231 & 1.029 & 0.753 \\
 0.955 & -1.997 & 1.507 \\
 -1.265 & 1.447 & 0.982 \\
\end{array}
\right)\quad
                (b_{ij})\simeq 
                \left(
\begin{array}{rrr}
 1.685 & 1.397 & -0.842 \\
 0.877 & 0.777 & -1.451 \\
 -1.544 & 1.425 & 1.359 \\
\end{array}
\right)$
            \\ \hline
            VEVs, Value &$(v_1,v_2)\simeq (0.158,0.079)\;,\qquad \mathcal{V}(\mathcal{Q})\simeq -0.488$\\ \hline
\end{longtable}
    
\bibliographystyle{unsrt}

\end{document}